\documentclass[a4paper,11pt]{article}
\usepackage{jcappub}
\usepackage{lineno}

\usepackage{soul}
\usepackage{amsmath,amsthm,amssymb}
\usepackage{graphicx}
\usepackage{url,hyperref}

\usepackage[dvipsnames]{xcolor} 
\usepackage{verbatim}
\usepackage{subcaption}

\usepackage{setspace} 
\AtBeginDocument{\setstretch{1.125}}

\usepackage{outlines} 
\usepackage{tensor} 
\usepackage{csquotes} 

\usepackage{blindtext}
\usepackage{tcolorbox}

\hypersetup{colorlinks=true,linkcolor=blue,citecolor=magenta,filecolor=magenta,urlcolor=blue}

\numberwithin{equation}{section} 
\renewcommand{\theequation}{\arabic{section}.\arabic{equation}}

\DeclareUnicodeCharacter{2212}{-} 

\usepackage{etoolbox}

\makeatletter
\patchcmd{\frontmatter@abstract@produce}
  {\vskip200\p@\@plus1fil
   \penalty-200\relax
   \vskip-200\p@\@plus-1fil}
  {}
  {}
  {}
\makeatother

\title{Reissner-Nordstr{\"o}m black holes in de Sitter spacetime: 
bounds with quasinormal frequencies}

\author[a,b]{Anna Chrysostomou,}
\affiliation[a]{IP2I, Universit{\'e} Claude Bernard Lyon 1, UMR 5822, CNRS/IN2P3, 4 rue Enrico Fermi, 69622 Villeurbanne Cedex, France}
\emailAdd{chrysostomou@ip2i.in2p3.fr}
\affiliation[b]{Department of Physics, University of Johannesburg, PO Box 524, Auckland Park 2006, South Africa}

\author[a,b]{Alan S. Cornell,}
\emailAdd{acornell@uj.ac.za}

\author[a,b]{Aldo Deandrea,}
\emailAdd{deandrea@ip2i.in2p3.fr}

\author[b]{Hajar Noshad,}
\emailAdd{hnoshad@uj.ac.za}

\author[c]{Seong Chan Park}
\emailAdd{sc.park@yonsei.ac.kr}
\affiliation[c]{Department of Physics and IPAP, Yonsei University, Seoul 03722, Republic of Korea}

\abstract{Rich physics can be divined from charged black holes subjected to extremal conditions. When applied in conjunction with principles like Weak Cosmic Censorship, this naturally leads to constraints on the mass and charge of the black hole. However, more nuanced principles such as the Weak Gravity Conjecture (WGC) and the recently proposed Festina-Lente (FL) bound can provide, respectively, upper and lower bounds  on elementary charged particles. 
In this study, we examine the quasinormal modes (QNMs) exhibited by a massive scalar test field carrying an electric charge, oscillating in the outer region of the black hole. These modes are subjected to the constraints imposed by the FL and WGC bounds. Our analysis provides insight into the behaviour of QNMs, particularly in regions that approach the extremal conditions of the black hole. Notably, in these regimes, the stability of the modes becomes precarious, particularly in the presence of a positive cosmological constant.
The implications of our findings are far-reaching and significant. They extend from safeguarding the principles of cosmic censorship to addressing the structural stability of the black hole's interior. Our semi-classical analysis presents compelling evidence suggesting that Strong Cosmic Censorship may be violated for black holes that are in close proximity to extremality within the context of Reissner-Nordström-de Sitter (RNdS) geometries.
}

\begin{document}
\maketitle


\section{Introduction \label{sec:intro}}
 
\par In a universe devoid of matter, black holes emerge as the simplest solution to the Einstein field equations (EFEs). From the ``no-hair" conjecture, which originated in the 1960s \cite{Israel1966,Gravitation1973}, black holes can be entirely characterised by their mass $m_{\rm BH}$, charge $q_{\rm BH}$, and spin $a_{\rm BH}$ \cite{Gravitation1973}. Their constrained parameter space and purely-geometric nature have promoted black holes and their analogues into highly versatile laboratories \cite{Nature_BlackHoleDefinitions} in which theoretical conjectures within thermodynamics, cosmology, and high-energy particle physics and astrophysics can be scrutinised, as well as classical, semi-classical, and quantum gravity. A pertinent example for the latter cases lies in the Swampland programme \cite{Vafa2005_Swampland}, in which low-energy effective field theories (EFTs) can be constrained using heuristic arguments from black hole mechanics.

\subsection{Constraints from the swampland}
\par The Weak Gravity Conjecture (WGC) is a well-established result thereof, based on the consideration of the emission of elementary particles of mass $m$ and charge $q$ from black holes of mass $M$ and charge $Q$ in Minkowski space-time. The formulation of the conjecture is as follows. At (super)extremality, a black hole in asymptotically-flat space-time has a mass (greater than or) equal to its charge $M/|Q| \geq 1$ in natural units. Black hole decay occurs by means of the radiation of elementary particles. If such elementary particles have a mass-charge ratio $m/\vert q \vert >1$, then the black hole would radiate all of its mass before losing its charge, resulting in a ``black hole" remnant of zero mass but non-zero charge. Citing the logical inconsistency of such a result, the WGC requires the existence of an elementary particle whose ratio of mass to gauge field charge is less than one. We can write this as 
\begin{equation} \label{eq:WGCmass}
m < g_1 M_P \;,
\end{equation}
\noindent where $M_P$ is the Planck mass and $g_1$ is the $U(1)$ gauge coupling \cite{ArkaniHamed2006_WGCorigins}. Within this work, we take $U(1)$ to be electromagnetism, unless stated otherwise.

\par It is worth noting that the fundamental nature of black holes is qualitatively and quantitatively altered upon introducing a non-zero cosmological constant, $\Lambda$, into the Einstein-Hilbert-Maxwell theory. In the case of $\Lambda>0$, as considered in this work, it is immediately clear that the maximum possible mass of a black hole is reduced: during stellar collapse, for example, the gravitational attraction must compete against the repulsive vacuum energy. We shall see explicitly in section \ref{subsec:RNdSBH} how introducing $\Lambda$ decreases the extremal black hole mass-charge ratio from one to $M/Q \lesssim 0.9428$. More subtle concerns include the impact of the Gibbons-Hawking entropy \cite{GibbonsHawking1977_BHTherm} on the black hole system, particularly with respect to black hole decay (see Ref. \cite{MossNaritaka2021_KerrdSevaporation} and references therein), as well as the question of defining mass \cite{Dolan2018_DefiningMassdS} and black hole observables in de Sitter (dS) space-times \cite{Anninos2012_Musings}. Naturally, these have direct implications for any \textit{gedankenexperiment} conducted within the black hole laboratory. In this work, we confine our discussions to the $\Lambda>0$ case; specifically, we embed a static, electrically-charged black hole (i.e. the Reissner-Nordst{\"o}m (RN) solution) into asymptotically-dS space-time.

\par Of the maximally-symmetric EFE solutions, dS space-time best reflects our current dark-energy-dominated cosmological era. Two key experiments in 1998 \cite{SupernovaSearchTeam1998_LCDM,
SupernovaCosmologyProject1998_LCDM} provided the initial empirical evidence that the universe is expanding at an accelerated rate; surveys continue to support the existence of a positive cosmological constant \cite{Planck2018}, such that the $\Lambda$CDM model remains our standard model of cosmology. Within theory, interest in dS space-time is pervasive \cite{Anninos2012_Musings} due to its holographic dual description via conformal field theory \cite{Strominger2001_dScft}, the still-unresolved question of dS stability in quantised theories \cite{vanRiet2018_dSvacuua,Dine2020_dSobstacles}, and its generalisable thermodynamic properties \cite{DolanMann2013_dStherm}. 

\par This last point is particularly interesting. As a consequence of the exponential expansion associated with the positive vacuum energy, a black hole in a dS universe is surrounded by a cosmological horizon beyond which information becomes inaccessible. Hawking radiation  \cite{Hawking1974_HawkT,Hawking1975_HawkRad} emanates from this horizon, such that the cosmological horizon is associated with a dS temperature $T_{\rm dS}$ and Gibbons-Hawking entropy $S_{\rm dS}$ \cite{GibbonsHawking1977_BHTherm}. Except in specific cases (discussed in sections \ref{subsec:RNdSBH} and \ref{subsec:nearNariai}), the black hole temperature $T_{\rm BH}$ differs from $T_{\rm dS}$. For a sufficiently large black hole, however, the system drifts towards thermal equilibrium.

\par In the case of Reissner-Nordst{\"o}m de Sitter (RNdS) black holes, the transition towards equilibrium is mediated by an exchange of mass and charge between cosmological horizon and black hole. This process was recently investigated by Refs. \cite{vanRiet2019_FLevapBHdS,vanRiet2021_FL}, in an attempt to understand how RNdS black holes decay and to extend the principles of the WGC to dS space-times. There, the authors established the ``Festina-Lente" (FL) bound for elementary particles of mass $m$ and charge $q$ discharging the black hole,
\begin{equation} \label{eq:FLbound}
\frac{m^4}{8 \pi \alpha } \geq V \;,
\end{equation}
\noindent where $\alpha = g_1^2 q^2 / 4 \pi \sim 1/137$ is the fine structure constant, with $g_1$ as the $U(1)$ gauge coupling, and $V= \Lambda / 8 \pi G = 3 M^2_P H^2$, with $H^2=\Lambda/3$, as the gravitating vacuum energy. We can write this as $m^2 \geq \sqrt{6} g_1 q M_P H$. In natural units, the Planck scale is $M_P \sim 10^{27}$ eV and the current Hubble scale is $H \sim 10^{-33}$ eV. 

\par Since we consider the $U(1)$ to be electromagnetism, the scale is set at $\sqrt{g_1 M_P H} \sim 10^{-3}$ eV, around the vacuum energy density scale/neutrino mass scale.\footnote{Since neutrinos are electrically neutral, the FL bound does not apply. Attempts to extend the FL bound beyond fields charged under $U(1)$ are being explored, such as in Ref. \cite{SCParkDYCheong2022_FLmili}. As the electron mass is determined by the vacuum expectation value (VEV) of the Higgs, the shape of the Higgs potential is constrained by the FL bound as discussed in Ref.
\cite{Lee:2021cor}.} The bound is therefore the geometric mean between our current Hubble scale and the Planck scale (see Fig. 5 of Ref. \cite{vanRiet2019_FLevapBHdS}). For the lightest electrically-charged particle in the Standard Model (SM), the electron, the particle mass $m_e \sim 10^5$ eV comfortably satisfies the FL bound. On the other hand, the electron saturates the WGC by 19 orders of magnitude. 

\par Provided the charge carrier is sufficiently heavy, Eq. (\ref{eq:FLbound}) demonstrates that the RNdS black hole will evaporate to dS space-time, in the usual fashion \cite{Anninos2012_Musings}. However, in the case of very light particles (i.e. $m^2 \ll \sqrt{6} g_1 q M_P H$) discharging from very large charged black holes, such that $r_{\rm BH} \sim r_{\rm dS}$, the charge depletion is near-instantaneous and the result is a Big Crunch solution. In other words, the black hole passes from sub-extremal to super-extremal, leading to a naked curvature singularity, which is a violation of cosmic censorship \cite{Penrose1964_CC1,Penrose1969_CC2,Hawking1970_CC3}. We shall return to this point in a moment.

\par As shown in Ref. \cite{vanRiet2021_FL}, we can combine Eqs. (\ref{eq:WGCmass}) and (\ref{eq:FLbound}) to determine an upper and lower limit, respectively, on the mass of the elementary particle being discharged from a RNdS black hole,
\begin{align} \label{eq:FullBound}
\sqrt{8 \pi \alpha V} \sim \sqrt{6} g_1 q M_P H & < m^2 <  2 g^2_1 q^2 M^2_P \sim 8 \pi \alpha M_P^2 \;.
\end{align}
\noindent This corresponds to $10^{-3} \; \text{eV} \lesssim m \lesssim  10^{26} \; \text{eV}$ in the case of a $U(1)$ charge in our current universe. For consistency, the WGC-based limit is derived under the assumption that $r_{\rm BH} \ll r_{\rm dS}$ and that cosmic censorship must be preserved \cite{vanRiet2021_FL}. In contrast, the FL bound is derived for black holes of size $r_{\rm BH} \sim r_{\rm dS}$.  We discuss this further in section \ref{subsec:nearNariai}.

\par Eq. (\ref{eq:FullBound}) is a strong indication of the value of studying the RNdS black hole within the broader context of well-established principles like cosmic censorship, and under the assumption of black hole stability. In the case of a RNdS space-time, these ideas are interlinked and can be investigated through the study of the perturbations of the black hole in question. 

\subsection{Cosmic censorship and black hole stability in RNdS space-times}

\par Cosmic censorship is closely related to the causal structure of a black hole. In the classical theory of general relativity (GR), a curvature singularity lurks at the centre of a black hole while a coordinate singularity is associated with the event horizon. Since the curvature singularity is irredeemably ill-defined within GR, it represents a failure of the EFEs. To preserve the predictive power of GR, this necessitates the presence of an event horizon capable of hiding the naked singularity from observers \cite{Penrose1964_CC1,Penrose1969_CC2,Hawking1970_CC3}. The Weak Cosmic Censorship (WCC) conjecture requires only that this curvature singularity is hidden from distant observers by an event horizon stable against perturbations \cite{Penrose1969_CC2,Wald1997_WCC}. The Strong Cosmic Censorship (SCC) conjecture, on the other hand, specifies that the singularities within black hole interiors must be space-like, that they appear only on space-like or null (light-like) surfaces, such that the evolution of initial data can be uniquely predicted (see Ref. \cite{Chambers1997_SCC} for a review). SCC becomes a contentious issue in RN black holes characterised by an inner Cauchy horizon within which surfaces of constant $r$ $-$ and thus the curvature singularity $-$ are time-like \cite{Chandrasekhar1983}. This marks a breakdown in GR's deterministic nature: an observer crossing the Cauchy horizon on a time-like trajectory will enter a space-time in which past-directed null geodesics terminate in the time-like singularity; the evolution of initial data is therefore subject to yet-unknown boundary conditions at the singularity and thus cannot be guaranteed to be unique.  

\par Though the WCC conjecture is satisfied (i.e. the singularity is encircled by the event horizon and thus covered), SCC appears violated. Whether this is indeed the case has been argued extensively. Here, however, we address the original argument put forth by Penrose, namely that the inner horizon is highly unstable against ingoing perturbations \cite{Penrose1999_CC}. Ingoing radiation entering the event horizon becomes infinitely blue-shifted at the Cauchy horizon, accumulates along this inner horizon, and causes the curvature to diverge, thereby transforming the Cauchy horizon into a singularity.\footnote{Initially, the singularity that forms is light-like. Over time, the singularity becomes space-like \cite{refHorowitzBook}. As discussed in Refs. \cite{PoissonIsrael1990_BHinterior,Dafermos2003_RNinterior}, the perturbed Cauchy horizon becomes a mass inflation singularity that is strong enough to impose the breakdown of the EFEs.} This exponential divergence is governed by $\kappa_-$, the surface gravity \cite{BardeenCarterHawking1973_BHmechanics} of the Cauchy horizon. Since the EFEs then cannot be continued past the Cauchy horizon, determinism is re-established and SCC is rescued.

\par However, this argument does not take into the account the behaviour of perturbations in the exterior of the black hole, which compete against this blue-shift effect \cite{Dafermos2003_RNinterior,Dafermos:2012np}. In asymptotically-dS space-times, this competition is compounded by the accelerating cosmic expansion introduced by the vacuum energy \cite{Dafermos2018_BlueShiftLambda}. As the validity of SCC relies upon the blue-shift mechanism, the ``red-shift" driven by the cosmological constant could lead to the exterior behaviour overwhelming the interior. 

\par This has been the subject of much discussion in recent years e.g. Refs. \cite{Cardoso2017_QNMsSCC,Cardoso2018_QNMsSCC,Hod2018_SCC,
Mo2018_SCC,DiasReallSantos2018_SCC,DiasEperonReallSantos2018_SCC,DiasReallSantos2018_SCC_Rough,GimGwak2019_RNdS-Lyapunov,Konoplya2022_SCC}. These works centre on Refs. \cite{Hintz2015_SCC,Costa2016_SCC}, where it was shown that for a non-degenerate RNdS black hole of dimension $d \geq 4$, there exists some $\beta>0$, dependent only on the black hole parameters, such that the exponential decay for massive and neutral scalar fields is governed by the expression
\begin{equation} \label{eq:Hintz}
\vert \Phi  \vert \leq C e^{-\beta \kappa_- t} \;, \quad \beta \equiv -\frac{\mathbb{I}m \{ \omega^{n=0} \}}{\vert \kappa_- \vert} \;, 
\end{equation}  
\noindent for small mass $m>0$. $\Phi$ is a linear scalar perturbation, $C\geq 0$, and $\mathbb{I}m \{ \omega^{n=0} \}$ is related to the inverse damping rate of the longest-lived quasinormal mode (QNM)\footnote{As shall be discussed further in section \ref{subsec:potential}, QNMs are damped proper modes characterising the late-time evolution of a perturbed object. For RNdS black hole, they obey boundary conditions that are purely ingoing (outgoing) at the event (de Sitter) horizon. We label QNMs by the monotonically-increasing overtone number $n$.} (see Refs. \cite{Cardoso2017_QNMsSCC,DiasReallSantos2018_SCC} for further discussion). Although originally conjectured for electrically-neutral fields, appendix A of Ref. \cite{Cardoso2018_QNMsSCC} demonstrated that the stability of the Cauchy horizon still depends on $\beta$ for $q \neq 0$. It is from this requirement that the criterion for the preservation of SCC,
 \begin{equation} \label{eq:SCCvalid}
     SCC \iff \beta < \frac{1}{2} \;,
 \end{equation}
 \noindent can be derived. There are a number of instances recorded in the literature in which Eq. (\ref{eq:SCCvalid}) is not respected. For massless, electrically-neutral scalar fields within near-extremal RNdS black holes, Ref. \cite{Cardoso2017_QNMsSCC} found that $1/2 < \beta < 1$. In Ref. \cite{DiasReallSantos2018_SCC}, large scalar field charge resulted in a violation of the SCC, which was confirmed in Ref. \cite{Cardoso2018_QNMsSCC} for a fixed value of the cosmological constant. In the case of gravito-electromagnetic QNMs, it was shown in Ref. \cite{DiasReallSantos2018_SCC_Rough} that the Christodoulou and $C^2$ formulations of SCC are always violated in the parameter space close to extremality. For sufficiently large black holes, this violation of SCC extends to $C^r \; \forall \; r$. We wish to explore this interplay between black hole and scalar field parameters in greater detail, guided by the WCC conjecture, the WGC, and the FL bound.
 
\par However, it is important to recognise that any exploration of the properties of the black hole space-time assumes the (classical) stability of the probed black hole and the perturbations thereof. Black hole stability analyses through the study of linearised perturbations were initiated in the 1957 investigation of the asymptotically-flat Schwarzschild space-time by Regge and Wheeler \cite{refRW}, and have been extended to a wide variety of contexts (see Refs. \cite{refBertiCardoso,refKonoplyaZhidenkoReview} for historical overviews and further discussion). For QNMs with a harmonic time dependence $e^{-i \omega t}$, the criterion for stability is a decaying mode, such that $\mathbb{I}m \{ \omega \} < 0 $. However, if there is at least one growing mode, it is understood that the space-time is unstable, where the instability growth rate is proportional to the imaginary part of the growing QNM \cite{refKonoplyaZhidenkoReview}. 

\par In the case of the RNdS black hole space-time, stability was well-established for massless, uncharged fields $ d \leq  6 $ \cite{refIKrn,refIKchap6,refKonoplya2009}. When charged, however, instabilities may arise. This was first reported in Ref. \cite{ZhuZhang2014_RNdSinstability} for massless QNMs with an angular momentum number of zero, where superradiance was considered to be the cause. Superradiance, in this case, refers to a phenomenon observed in scattering problems in which the electromagnetic energy of a black hole is extracted and radiated away by an ingoing wave; the reflected wave is characterised by an amplitude larger than the incident wave \cite{Bekenstein1973_SR}.\footnote{Superradiance within the black hole community has recently garnered a great deal of attention for its applications in exploring the mass limits of ultralight bosons, novel space-time geometries in dimensions $d \geq 4$, and the gauge/gravity duality (see Ref. \cite{BritoCardosoPani2020_Superradiance} and references within).} The coupling between the particle charge and the electromagnetic field of the black hole overcomes the gravitational attraction between the particle and black hole mass, $qQ \gg \mu M$ \cite{KonoplyaZhidenko2014_RNdSrprc}. This phenomenon has been observed and re-examined by several authors in the RNdS context \cite{KonoplyaZhidenko2013_dRNdSinstability,
KonoplyaZhidenko2014_RNdSrprc,
ZhuZhang2014_RNdSinstability,
DiasReallSantos2018_SCC,
DiasSantos2020_RNdSinstability,
Papantonopoulos2022}. For the parameters considered in these references, all unstable modes met the criteria for superradiance; all non-superradiant modes were found to be stable. As such, it was concluded in Ref. \cite{KonoplyaZhidenko2013_dRNdSinstability} that supperradiance serves as a necessary (not sufficient) condition for instability. We explore some of the implications for this in section \ref{subsec:superrad}.

\par Noticeably, black hole stability and the preservation of SCC for charged black holes depends on how effectively the exterior damps perturbations. This information can be extracted directly from QNM analyses. Our focus in this work is therefore centred on understanding the behaviour of a charged, massive, minimally-coupled scalar test-field in the exterior region of the RNdS black hole space-time for the full available parameter space, as defined from considerations of the WCC for the RNdS phase space as well as the WGC and FL bound for the mass of the perturbing field. Like in Ref. \cite{vanRiet2019_FLevapBHdS}, we focus on a single species of charged particle: here, a massive scalar field with a $U(1)$ charge $q$ and mass $\mu$. The mass-charge regimes of interest include the FL-WGC bound of $10^{-3}$ eV $\leq m \leq 10^{26}$ eV; the regime corresponding to instability, $qQ \gg \mu M$ \cite{KonoplyaZhidenko2014_RNdSrprc}; the regime corresponding to the strongest superradiant instabilities for astrophysical black holes \cite{BertiBrito2019_UltraLight1GW,BertiBrito2017b_UltraLightLIGOLISA}, $\mu M \sim 1$. We begin with an overview of the 4D RNdS phase space, with a specific focus on the $r_{\rm BH} \sim r_{\rm dS}$ limit, followed by a description of the QNM formalism for massive, charged scalar perturbations oscillating within the RNdS space-time. We then proceed in section \ref{sec:QNFs} to a discussion of some known results from the literature that connect QNMs to the principles of the WGC and SCC. We comment on how our results within the phase space relate to these known issues in the literature. 

\section{The RNdS phase space: a ``sharkfin" and the potential within}

\par The RNdS black hole space-time has been studied extensively, with a particular interest in black hole pair creation, thermodynamics, and decay (see, for example, Refs. \cite{Romans1991_ColdLukewarmRNdS,
Mann1995_ChargedBHpairs,
Bousso1996_Nariai,
refCardosoDiasLemos,
CasalsDolan2009_Nariai,
Belgiorno2009_chargedBHs,
Belgiorno2010_chargedBHs,
AntoniadisBenakli2020_WGCdS,
vanRiet2019_FLevapBHdS,
vanRiet2021_FL}). In this section, we shall review the main features of the solution space of the 4D RNdS black hole space-time that precludes a naked singularity (where we refer the reader to Ref. \cite{AntoniadisBenakli2020_WGCdS} in particular for further details in $d=4$ and appendix A of Refs. \cite{refIKrn,refNatarioSchiappa} for the $d$-dimensional case). Thereafter, we shall discuss the behaviour of the QNM effective potential for a scalar test field of non-zero mass and charge with respect to the RNdS phase space. 

\subsection{A review of the RNdS black hole solution space \label{subsec:RNdSBH}}

\par We begin with a stable (3$+$1)-dimensional dS space-time whose lifetime is approximately of the order of the Hubble scale (and not exceeding $(1/H) \log H$ to satisfy the trans-Planckian censorship conjecture \cite{BedroyaVafa2019_TCC}). In geometric units $\hbar = c = k_B = 1$ and using the $(-+++)$ signature, we consider the Einstein-Hilbert-Maxwell action in an empty dS space-time as
\begin{equation} \label{eq:action}
S =  \int d^4x \sqrt{-g} \left[ \frac{1}{2 \kappa^2}\left(R - 2\Lambda  \right) - \frac{1}{4g_1^2} F^{\mu \nu} F_{\mu \nu} \right] \;.
\end{equation}  
\noindent Here, $\kappa^2~=~8 \pi G~=~M^{-2}_P$ relates the gravitational coupling $\kappa$ to Newton's gravitational constant $G$ and the Planck mass $M_P$. The dS radius $L_{dS}$, Hubble parameter $H$, and cosmological constant $\Lambda >0$ are related by $H^2~=~L^{-2}_{dS}~=~\Lambda /3$. The geometry of the space-time is encoded in the Ricci scalar curvature $R~=~g^{\mu \nu} R_{\mu \nu}$ and the metric $g~=~\det |g_{\mu \nu}|$; the metric tensor $g_{\mu \nu}$ is in turn defined in terms of the characteristic black hole parameters, mass $m_{_{_{\rm BH}}}$ and charge $q_{_{_{\rm BH}}}$ per the `no-hair conjecture' \cite{Gravitation1973}, as well as $\Lambda$.\footnote{Recall that $\Lambda$ is a parameter of the `space of theories', a model-dependent degree of freedom, rather than a black hole parameter. In the $\Lambda$CDM model, it is understood that $\Lambda$ is associated with the vacuum energy of the scalar field that once drove inflation. In other words, the potential of the inflaton acts as an effective cosmological constant whose value is near-zero in our present epoch \cite{Bousso1996_Nariai}.} For the electromagnetic field strength tensor $F^{\mu \nu}$, $g_1$ is the $U(1)$ gauge coupling. Since we are concerned only with the electrically-charged RNdS case, the non-zero component of $F^{\mu \nu}$ is
\begin{equation}
F_{tr} = \frac{g_1^2}{4 \pi} \frac{q_{_{_{\rm BH}}}}{r} dt \wedge dr\;,
\end{equation}
with a purely electric gauge potential 
\begin{equation} \label{eq:vecA}
A = \Phi dt \;, \quad \Phi = \frac{g_1^2}{4 \pi} \frac{q_{_{_{\rm BH}}}}{r} \;.
\end{equation}

\par The Lagrangian of Eq. (\ref{eq:action}) admits the static and spherically-symmetric black hole solution \cite{Kramer2003_ExactEFEsols} 
\begin{equation} \label{eq:metric}
ds^2 = -f(r) dt^2 \; + \; f(r)^{-1} dr^2 \; + \; r^2 \left(d\theta^2 +  \sin^2 \theta  d \phi^2 \right) \;, 
\end{equation}
\noindent written in terms of the usual Schwarzschild coordinates $(t,r,\theta,\phi)$, with $t \in (-\infty,+\infty)$, $\theta \in (0,\pi),$ and $\phi \in (0,2\pi)$. Surfaces of constant $r$ are space-like in the region exterior to the black hole event horizon, $r \in (r_+,r_c)$, bounded by the event horizon and the cosmological horizon, respectively. In this region, $t$ is time-like as $f(r)>0$. Using natural units, we can express this metric function as

\begin{equation} \label{eq:ffull}
f(r) = 1 - \frac{2Gm_{_{\rm BH}}}{r} + \frac{Gg^2_1 q_{_{\rm BH}}}{4 \pi r^2} - \frac{\Lambda}{3} r^2 \;.
\end{equation}
\noindent While the physical quantities $m_{_{\rm BH}}$ and $g_1 q_{_{\rm BH}}$ lend themselves to characteristic black hole length scales \cite{Bekenstein2003_BHinfo}, all three parameters have dimensions of length \cite{AntoniadisBenakli2020_WGCdS}:
\begin{align}
Gm_{_{\rm BH}} = \frac{\kappa^2}{8 \pi}m_{_{\rm BH}} &= M \;, \\
\frac{G}{4 \pi r^2}g^2_1 q_{_{\rm BH}} = \frac{\kappa^2}{32 \pi^2 r^2}g^2_1 q_{_{\rm BH}} & =  Q^2 \;, \\
\Lambda = \frac{3}{L^2_{dS}} \;.
\end{align}
\noindent Note that under this set of units\footnote{In the case of Planck units ($\hbar=c=G=1$), the fundamental unit of electric charge is given in terms of the fine structure constant $\alpha$, with $e \sim \sqrt{1/137} \sim 0.1 M_P \sim 10^{-6}$ grams \cite{refHorowitzChap1}.}, we can compare mass and charge directly
\begin{equation} \label{eq:mqratio}
\frac{M^2}{Q^2} = \frac{\kappa^2}{2} \frac{m_{_{\rm BH}}^2}{g^2_1 q^2_{_{\rm BH}}} \;.
\end{equation} 

\par With these parametrisations in place, we can express the metric function simply as
\begin{equation} \label{eq:f}
f(r) = 1 - \frac{2M}{r} + \frac{Q^2}{r^2} - \frac{r^2}{L^2_{dS}}  \;.
\end{equation}
\noindent The global structure of the black hole spacetime is determined by $(i)$ the behaviour at $r=0$ (where there is a scalar curvature singularity); $(ii)$ the behaviour at $r = \infty$; $(iii)$ the behaviour at the Killing horizons determined by $f(r)=0$. Specifically, the roots of the metric function dictate the causal structure of the space-time, and depend strongly on the values of $M$, $Q$, and $L_{dS}$. When cosmic censorship is preserved, $f(r)=0$ has four real roots corresponding to the inner (Cauchy) horizon $r_-$, the outer (event) horizon $r_+$, the cosmological horizon $r_c$, and $r_0 = -(r_- + r_+ + r_c)$. The positive roots are the three Killing horizons of the space-time, with $r_- \leq r_+ \leq r_c \leq L_{dS}$.\footnote{The internal horizon serves as a light-like boundary, such that the geodesics within the horizon are time-like rather than space-like. Here, $f(r)<0$ such that $t$ is space-like and the gravitational attraction $-2M/r$ dominates. Similar behaviour is observed for $r>r_c$: $f(r)<0,$ albeit the cosmological constant term dominates over the attractive gravitational potential and the repulsive electromagnetic potential of the black hole \cite{refHorowitzChap1,AntoniadisBenakli2020_WGCdS}.} The metric function can also be written in terms of the horizons,
\begin{equation} \label{eq:fhor}
f(r) = \frac{(r-r_-)(r-r_+)(r_c-r)(r-r_0)}{r^2 L^2_{dS}} \;.
\end{equation}
\par
\begin{figure}[t]
\centering
\includegraphics[width=0.8\linewidth]{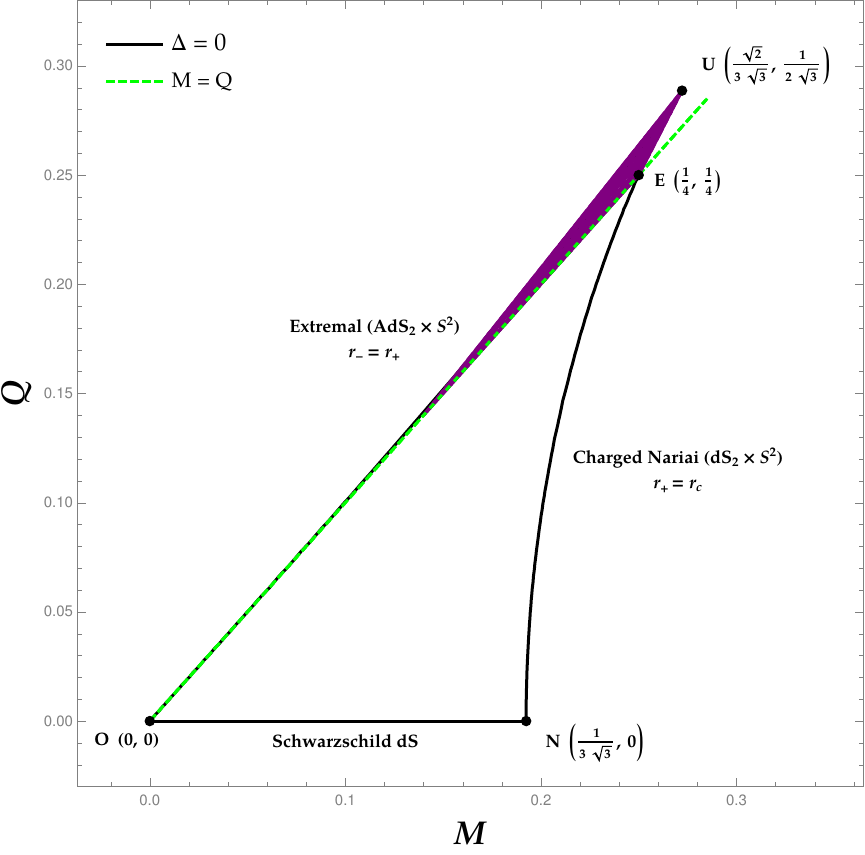}
\caption{ \textit{A two-dimensional projection of the parameter space for $H^2 = \Lambda/3 =1$ for the 4D RNdS black hole. The boundary of the RNdS phase space or ``sharkfin" corresponds to extremal conditions. At point $O$ lies pure dS space $(Q=M=0)$. Along the ON line, there is the Schwarzschild dS family of solutions $(Q=0, M>0),$ with the uncharged Nariai case at point $N$. The charged Nariai branch, at which $r_+ = r_c$, extends along NU; the extremal branch of ``cold" solutions, with $r_- = r_+$, follows OU. On this branch, $Q \sim Q_{ext} \sim r_+$, with $Q_{ext}$ defined in Eq. (\ref{eq:qext}). These branches terminate in the ``ultracold" solution, where $r_- = r_+ = r_c$; Hawking temperature goes to zero and the local geometry is $\mathbb{M}_2 \times \mathbb{S}^2$. 
}}
\label{fig:sharkfin}
\end{figure}

 In Fig. \ref{fig:sharkfin}, we sketch the parameter space of a RNdS black hole, provided with great detail in Refs. \cite{Bousso1996_Nariai,AntoniadisBenakli2020_WGCdS,
Benakli2021_DilatonicAdS}. Here, we introduce the polynomial
\begin{equation}
\Pi(r) \equiv -r^2 f(r) = -r^2 +2Mr -Q^2 + L^2_{dS} r^4 \;.
\end{equation}
\noindent The discriminant thereof is 
\begin{equation} 
\Delta \equiv -16 L^{-2}_{dS} \left[27 M^4 L^{-2}_{dS} - M^2(1 + 36Q^2L^{-2}_{dS}) + (Q + 4Q^3L^{-2}_{dS})^2 \right] \;.
\end{equation}
\noindent 
We set $L_{dS}=1$ and plot $\Delta = 0$, thereby obtaining the phase diagram or ``sharkfin" provided in Fig. \ref{fig:sharkfin}. 

\par If we set $\Delta = 0$ and solve for $M$ in terms of $\Lambda$ and $Q$, we reproduce the known bound on $M^2 \Lambda$ \cite{Romans1991_ColdLukewarmRNdS,
Bousso1996_Nariai},
\begin{equation} \label{eq:Mlimit}
M^2 \Lambda \leq \frac{1}{18} \left[ 1 + 12 Q^2 \Lambda + (1 - 4 Q^2 \Lambda)^{3/2}  \right] \;.
\end{equation}
\noindent We deduce then that $M^2 \Lambda$ has an upper bound of $1/9$. In Fig. \ref{fig:sharkfin}, where we parametrise $L^{-2}_{dS} = \Lambda/3 =1$. $M=\sqrt{2/27}$ is marked on the tip of the sharkfin as Point $U$. Traditionally \cite{Romans1991_ColdLukewarmRNdS,Bousso1996_Nariai}, the constraint on the black hole charge in 4D comes from the Bogomoln'yi bound for small values of $M^2 \Lambda$ \cite{Romans1991_ColdLukewarmRNdS},
\begin{equation} \label{eq:Qlimit}
\frac{Q^2}{M^2} \lesssim 1 + \frac{1}{3} (M^2 \Lambda) + \frac{4}{9} (M^2 \Lambda)^2 + \frac{8}{9} (M^2 \Lambda)^3 + \mathcal{O}(M^8 \Lambda^4) \;.
\end{equation} 
\noindent For $\Lambda = 3$ and $M=\sqrt{2/27}$, Eq. (\ref{eq:Qlimit}) approximately produces the $Q = 1/\sqrt{12}$ result obtained by solving for $Q$ from $\Delta = 0$ and observed at the tip of the sharkfin. 
If we define for $L^2_{dS}=1$ and $\underline{\Delta} = -\Delta /16$
then the corresponding horizons are given by
\begin{equation}
r_-  =  - a + b \;, \quad
r_+  =  + a - b \;, \quad
r_c  =  + a + b \;, \quad
r_0  =  - a - b \;. \label{eq:radii}
\end{equation}  
Here,
\begin{equation}
a=\frac{1}{2 \sqrt{3}} \sqrt{\frac{(1 + X)^2-12 Q^2}{X}}\,,  \quad \quad 
b=\frac{1}{2} \sqrt{\frac{4}{3} - \frac{1 - 12 Q^2}{3X} - \frac{X}{3} + \frac{2M}{a}} \;,
\end{equation}
and 
\begin{equation}
X = \left(-1 + 54M^2 -36Q^2 - 2\sqrt{27} \sqrt{ \underline{\Delta}} \right)^{1/3} \;.
\end{equation} 

\noindent At each horizon, we can calculate a surface gravity $\kappa_i$ and a Hawking temperature $T_i$ \cite{Hawking1974_HawkT} for $i \in \{-,+,C\}$,
\begin{equation}
\kappa_i = \frac{1}{2}  \frac{d}{dr} f(r) \bigg \vert_{r=r_i} \;, \quad T_i = \frac{\kappa_i}{2 \pi} \;.
\end{equation}
\noindent These assume a non-extremal black hole, where $\kappa_c <0$ and $|\kappa_-| > |\kappa_+|$ \cite{MossMyers1998_CC}. This is evidenced using Eq. (\ref{eq:fhor}),
\begin{align}
\kappa_- & =  - \frac{(r_+ - r_-)(r_c - r_-)(r_- - r_0)}{2r_-^2 L^2_{dS}} \label{eq:km} \\
\kappa_+ & =  + \frac{(r_+ - r_-)(r_c - r_+)(r_+ - r_0)}{2r_+^2 L^2_{dS}} \label{eq:kp}  \\
\kappa_c & =  -\frac{(r_c - r_-)(r_c - r_+)(r_c - r_0)}{2r_c^2 L^2_{dS}} \;.\label{eq:kc} 
\end{align}

\par For an extremal black hole, $f(r)=0$ has a double zero such that the horizon is degenerate and the surface gravity vanishes \cite{refHorowitzChap1}. For an extremal black hole, $\kappa_- = \kappa_+ = 0$; it was shown in Ref. \cite{DiasReallSantos2018_SCC} that this occurs when 
\begin{equation} \label{eq:qext}
Q = Q_{ext} \equiv y_+ r_c \sqrt{\frac{1 + 2y_+}{1+2y_+ + 3y_+^2}} \;, \quad y_+ = \frac{r_+}{r_c} \;.
\end{equation}
\noindent 
We see that $Q_{ext} \sim r_+$ (as suggested by Ref. \cite{DiasReallSantos2018_SCC} for the extremal RNdS black hole) on the $OU$ line, with closer agreement for lower values of $Q$. We provide the \href{https://github.com/anna-chrys/RNdS_QNMs}{interactive Mathematica notebook} for numerical confirmation of this.

\par We note then that for a non-extremal black hole, the horizons $r=r_+$ and $r=r_{C}$ are generally not in thermal equilibrium; the Hawking radiation coming from each horizon generally does so at different temperatures.  However, there are two ``families" of RNdS solutions for which the black hole and cosmological horizon temperatures do match: the ``lukewarm” black hole family ($M = Q$) and the charged Nariai black hole family ($r_+ = r_c$)  \cite{Zhang2016_ThermRNdS}.

\par In flat space-time, setting $M=Q$ leads to an extremal black hole. In dS space-time, however, the situation is more complicated. Thermal radiation isotropically pervades dS space-time \cite{GibbonsHawking1977_BHTherm}; the black hole evolves towards thermal equilibrium such that its final configuration is achieved when the event horizon reaches the same temperature as that of the surrounding bath i.e. when $T_+$ is equivalent to that of the dS edge \cite{Romans1991_ColdLukewarmRNdS}. As such, $\kappa_+ = \kappa_c$. The $M=Q$ line is demarcated on the phase diagram of Fig. \ref{fig:sharkfin}. 

\par We can then divide the phase diagram of Fig. \ref{fig:sharkfin} into two regions \cite{Romans1991_ColdLukewarmRNdS,
Brill1993_RNdSextrema,
Mann1995_ChargedBHpairs,
Bousso1999_QuantumStructredS}:
\begin{outline}
\1[$(i)$] The shaded $Q^2 > M^2$ region above the $M = Q$ line, where the RNdS black hole is colder than the lukewarm solution. Since $T_c <0$, the black hole absorbs radiation from the cosmological horizon. The region is bounded by the $OU$ line: as $Q^2$ increases with respect to $M^2$, the inner and outer black hole horizons eventually coincide,  leading to the extremal ``cold" RNdS black hole condition. Note how the cosmological constant raises the upper-bound of the black hole charge-mass ratio: in asymptotically-flat space-time, $Q/M \leq 1$ and $Q>M$ indicates super-extremality; in asymptotically-dS space-time, $Q/M \leq 1.06066.$
\1[$(ii)$] Below the lukewarm line is the $Q^2 < M^2$ region, where the black hole will be stabilised by the Gibbons-Hawking radiation from the cosmological horizon. The RNdS black hole cannot become arbitrarily cold. The region is bounded by the $NU$ line, upon which $r_+ = r_c$. The outer and cosmological horizons are in thermal equilibrium, leading to the charged Nariai solution where the mass is maximal at a given charge \cite{Bousso1996_Nariai}. 
\end{outline}
\par There is an important subtlety to address: a solution is ``cold" when the Hawking temperature vanishes \cite{Romans1991_ColdLukewarmRNdS}. On the extremal $OU$ branch, the inner and outer horizons coalesce and the surface gravity vanishes. However, on the charged Nariai branch, where $r_+ = r_c$, thermal equilibrium is reached. As discussed in Refs. \cite{vanRiet2019_FLevapBHdS}, a coordinate transform can be introduced to demonstrate that the two horizons approach one another but do not actually collide from the point of view of the geodesic observer as we approach
the charged Nariai branch. This is then a near-horizon limit.

\par As explained in appendix A of Refs. \cite{refNatarioSchiappa,refIKrn}, we can examine extremal conditions with respect to the black hole charge-mass relationship, and the useful constraint on $L_{dS}$,
\begin{equation}
    0 < L^2_{dS_{\pm}} =  \frac{1}{2}  \frac{(3M \pm \sqrt{9M^2-8Q^2})^3}{M \pm \sqrt{9M^2-8Q^2}} \;.
\end{equation}
Here, we shall observe how this corresponds to the region within the sharkfin diagram. 

\begin{outline}
    \1[$(i)$] For $Q^2 > M^2$, above the $M=Q$ line, we have a number of extremising configurations: 
    \2[$(a)$] We can impose extremal conditions on $Q$, such that $Q^2 = 9M^2/8$. This yields the ``RN-type" extremal condition that extremises the cosmological constant at the upper value such that $L^2_{dS} = L^2_{dS_{+}}$ i.e. $\Lambda~=~2/9M^2$. This yields the ``ultra-extremal" case  \cite{Romans1991_ColdLukewarmRNdS,Brill1993_RNdSextrema} at Point $U$ where there is only one real positive root,
\begin{equation} \label{eq:ultracold}
r_- = r_+ = r_c = \frac{3M}{2} \;.
\end{equation}
  \2[$(b)$] Alternatively, we can extremise $\Lambda$. For $\lambda = \lambda_-$, we have the ``RN-type'' extremal condition that gives two positive roots: one at $r=r_c$ and a degenerate horizon at
\begin{equation} 
r_- = r_+ = \frac{3M - \sqrt{9M^2-8Q^2}}{2}  \;. 
\end{equation}
This is the ``cold” black-hole case \cite{Romans1991_ColdLukewarmRNdS,Brill1993_RNdSextrema} on line $OU$.
\1[$(ii)$] For $Q^2 \leq M^2$, under the $M=Q$ line, the ``dS-type extremal condition" $L^2_{dS} = L^2_{dS_{+}}$ gives two positive roots,  
\begin{equation} 
r_+ = r_c = \frac{3M + \sqrt{9M^2-8Q^2}}{2}  \;. \end{equation}
\end{outline}
This is the ``extreme" or ``marginal" naked-singularity case, falling on the charged Nariai branch of $NU$ \cite{Romans1991_ColdLukewarmRNdS,Brill1993_RNdSextrema}

\subsection{The Nariai solution and the Festina-Lente bound \label{subsec:nearNariai}}

\par Solutions to the Einstein field equations (EFEs) can be determined through global analysis and topological methods, and have led to the establishment of mathematical frameworks describing the likes of astrophysical black holes (the Kerr solution) and the expanding universe (the Friedmann-Lema{\^i}tre-Robertson-Walker solution). However, not all solutions\footnote{For a compendium of known solutions to the EFEs, see Ref. \cite{Kramer2003_ExactEFEsols}.} have clear physical interpretations. Examples like the NUT-Taub and the Nariai solutions serve instead as fertile testing grounds for the development of physical intuition and numerical tools. Here, we shall consider the Nariai solution more closely.

\par The (uncharged) Nariai solution was constructed in 1950 \cite{Nariai1950_StaticSol,Nariai1951_NewSol} from the isotropic form of the line element,
\begin{equation}
ds^2 = -e^{\nu(r)} dt^2 + e^{\mu(r)} (dr^2 + r^2 d\theta^2 + r^2 \sin^2 \theta d \phi^2 ) \;.
\end{equation}
\noindent Assuming a homogenous static universe with spherical symmetry, Nariai determined that
\begin{equation} \label{eq:Nariai1}
ds^2 = \frac{1}{\Lambda} \left[ -(A \cos \{\log r\} + B \sin \{\log r \})^2 dt^2 + \frac{1}{r^2} (dr^2 + r^2 d\theta^2 + r^2 \sin^2 \theta d\phi^2 ) \right] \;,
\end{equation}
\noindent for arbitrary constants $A$ and $B$, satisfied the EFEs for an empty universe with a non-zero cosmological constant. We can reformulate this into a more tractable expression with the introduction of a few transformations \cite{Nariai1951_NewSol}, \textit{viz.} 
\begin{align}
t &= \tau \left( \frac{\Lambda}{A^2 + B^2} \right)^{1/2} \;, \nonumber \\
r & =  \exp \bigg \{ \pm \chi + \tan^{-1} \Bigg \{\frac{B}{A} \Bigg \} \bigg \} \;, \\
r_1 & =  L \sin \{ \chi \} \;, \nonumber
\end{align}
\noindent where $L^2=1 / \Lambda$. Upon setting these transformations into Eq. (\ref{eq:Nariai1}), we obtain 
\begin{equation}
ds^2 = -\cos^2 \chi d\tau^2 + L^2 (d\chi^2 + d\Omega^2) \;,
\end{equation}
\noindent for $d\Omega^2 = d\theta^2 + \sin^2 \theta d\phi^2$. Since $\sin^2 \chi = r_1/L$, we can rewrite this expression as
\begin{equation} \label{eq:Nariai2}
ds^2 = -\left( 1 - \frac{r_1^2}{L^2} \right) d\tau^2 + \left( 1 - \frac{r_1^2}{L^2} \right)^{-1} dr_1^2 + L^2 d\Omega^2 \;.
\end{equation}
\noindent We can compare this to the purely dS solution, where $L^2_{dS}=3/\Lambda$,
\begin{align}
ds^2 &= -\cos^2 \chi d\tau^2 + L_{dS}^2 (d\chi^2 + \sin^2 \chi d\Omega^2) \;,
 \\
&= -\left( 1 - \frac{r_1^2}{L_{dS}^2} \right) d\tau^2 + \left( 1 - \frac{r_1^2}{L_{dS}^2} \right)^{-1} dr_1^2 + r_1^2 d\Omega^2 \;. 
\end{align}
\noindent For the purely radial case $(d\Omega^2=0)$, Eq. (\ref{eq:Nariai2}) is nearly identical to pure dS space.  The Nariai space-time is spherically-symmetric, homogeneous and locally static; it is not isotropic or globally static. It has the geometry $dS_2 \times \mathbb{S}^2$ and a topology $\mathbb{R}\times\mathbb{S}^1\times\mathbb{S}^2$. The space-time satisfies $R_{\mu \nu} = \Lambda g_{\mu \nu}$, where $\Lambda=1/L^2$, and has constant Ricci scalar curvature, $R = 4\Lambda$. Furthermore, the space-time is symmetric $R_{\mu\nu\rho\sigma;\tau}=0$ (see also Refs. \cite{Bousso1996_Nariai,refCardosoLemos2003,
refCardosoDiasLemos,
CasalsDolan2009_Nariai} for further discussion).
 
\par A relationship between the Schwarzschild dS and Nariai solutions was first established by Ginsparg and Perry \cite{GinspargPerry1982_NariaidS} while studying the black hole thermodynamics of the latter: under a particular limiting procedure, the Nariai solution can be
generated as the event and cosmological horizons of the Schwarzschild dS approach one another.\footnote{Black holes in Minkowski and anti-dS space-times can be arbitrarily large. A static black hole in dS space-time, however, must fit within its cosmological horizon. For this reason, the Nariai solution is widely interpreted as the upper mass limit of a dS black hole: the event horizon grows such that the black hole occupies more and more of the dS space-time \cite{Cardoso2003_ExtremalSchw,
Molina2003_ExtremalDdimBHs,
Churilova2021_ExtremalAnal}. From Fig. \ref{fig:sharkfin}, we observe that introducing charge extends the upper bound on the black hole mass.} For this reason, the extremal Schwarzschild dS black hole, where $M = 1/\sqrt{9 \Lambda}$, is often referred to as the ``Nariai limit". With the inclusion of the Maxwell field, Bertotti \cite{Bertotti1959_BRsol} and Robinson \cite{Robinson1959_BRsol} introduced the charged Nariai solution. Then, analogously to Ref. \cite{GinspargPerry1982_NariaidS}, Hawking and Ross \cite{HawkingRoss1995_NariaiRNdS} obtained the charged Nariai solution as a limiting case of the RNdS black hole.

\par As explained in Ref. \cite{HawkingRoss1995_NariaiRNdS}, the limiting procedure itself relies upon an analytic continuation of the time coordinate into the complex plane $t \rightarrow i \tau$, which transforms the Lorentzian black hole into a Euclidean solution. For the consequent Euclidean metric to be positive definite, we require that $r \in (r_+,r_c)$ and $f(r)>0$. To ensure regularity on the horizon $r=r_i$, we identify $\tau$ with the period $\beta = 2 \pi/\kappa_i=1/T_i$. Conical singularities, on the other hand, can be removed from the metric if the temperature of the cosmological horizon and the black hole are identical. As specified in section \ref{subsec:RNdSBH}, this corresponds to the Nariai limit for the Schwarzschild dS case. For the RNdS black hole, the conical singularities at $r=r_+$ and $r=r_c$ can be removed in three ways: by setting $r_- = r_+$, $r_+=r_c$, or $|Q|=M$ \cite{Romans1991_ColdLukewarmRNdS}. 

\par Along the $NU$ branch of Fig. \ref{fig:sharkfin}, we consider the $r_+ = r_c$ case. Although we interpret this as a coalescing of the two horizons, there is a nuance to be highlighted: the proper distance between the two outer horizons remains finite in the $r_+ \rightarrow r_c$ limit, such that $r_+ \rightarrow \varrho - \epsilon$ and $r_c \rightarrow \varrho + \epsilon$ \cite{Romans1991_ColdLukewarmRNdS,
HawkingRoss1995_NariaiRNdS,
Anninos2012_Musings}. This is best illustrated by a change in coordinates, where we use the example from Ref. \cite{vanRiet2019_FLevapBHdS} that allows for a smooth transition of the black hole from a regular to extremal state. There, the coordinate $r=r_g$ is introduced, at which the competition between the gravitational attraction of the black hole and the accelerating expansion of the universe cancel and $f(r_g)=f'(r_g)=0$. A ``geodesic observer" situated at this point travels along a time-like Killing vector field which is also a geodesic. Suppose we let
\begin{equation}
\rho \rightarrow \frac{r-r_g}{\sqrt{ \vert f(r_g) \vert }} \;, \quad \tau \rightarrow \sqrt{ \vert f(r_g) \vert } \;t \;.
\end{equation}  
\noindent The metric then becomes
\begin{equation}
ds^2 = - \frac{f(r)}{\sqrt{ \vert f(r_g) \vert }} d\tau^2 + \frac{\sqrt{ \vert f(r_g) \vert }}{f(r)} d\rho^2 + r^2 d\Omega^2 \;,
\end{equation}
\noindent and the magnitude of the electric field is unchanged.  

\par From the perspective of the geodesic observer, the two horizons become infinitesimally close along the $NU$ branch, but they do not collide or overlap. There,
\begin{equation}
\frac{U(r)}{U(r_g)} \rightarrow 1-\frac{\rho^2}{L_{dS_2}^2}\;, \quad r^2 \rightarrow r_c^2 \;.
\end{equation}
\noindent This produces the $dS_2 \times S^2$ metric,
\begin{equation}
ds^2 = - \left( 1-\frac{\rho^2}{L_{dS_2}^2}\right)d\tau^2 + \left( 1-\frac{\rho^2}{L_{dS_2}^2} \right)^{-1} d \rho^2 + r_c^2 d \Omega^2 \;, 
\end{equation} where
\begin{equation}
L_{dS_2}^2 = \frac{2}{f''(r_c)} = \frac{1}{6} \left( \frac{1}{\sqrt{1-12Q^2}} + 1 \right) = \left(3 - \frac{Q^2}{r_c^4} \right)^{-1} \;.
\end{equation}

\noindent The corresponding $S^2$ radius on the $NU$ branch is 
\begin{equation}
r_c(Q) = \sqrt{ \frac{1}{6} \left( 1 + \sqrt{1-12Q^2} \right)} \;.
\end{equation}
\noindent We can see that this is equivalent to the value of the cosmological horizon $r_c$ in Eq. (\ref{eq:radii}) using the \href{https://github.com/anna-chrys/RNdS_QNMs}{interactive Mathematica notebook}.

\par Charged Nariai solutions represent a family of classically stable solutions within an extremal limit \cite{Romans1991_ColdLukewarmRNdS}. According to the WGC \cite{ArkaniHamed2006_WGCorigins}, ``extremal" (but not necessarily supersymmetric) black holes are expected to decay through the emission of massive charged particles whose ``elementary" electric and magnetic charge obey a mass-ratio $\mu/ |q| < M / \vert Q \vert$. In Minkowski space-time, where the WGC was proposed, $M / \vert Q \vert = 1$ (in appropriate units). When embedded in dS space-time, however, the evaporation of charged black holes is complicated by the exchange of mass and charge between the event horizon and the cosmological horizon. The study of this decay process led to the establishment of the FL bound in Ref. \cite{vanRiet2019_FLevapBHdS}.

\par The decay of the RNdS black hole is triggered by the Schwinger mechanism, whose decay rate is governed by  
\begin{equation} \label{eq:Schwinger}
\Gamma \sim \exp \bigg \{ - \frac{m^2}{qE} \bigg \} \;.
\end{equation}
\noindent The near-horizon electric field is of the order $E~\sim~\mathcal{O}(gM_PH)$, the usual electric field for a Nariai black hole.\footnote{At $Point \; U$ of Fig. \ref{fig:sharkfin}, $E=\sqrt{6}gM_PH$.} 

\par If $m \ll qgM_PH$, the electric field of the charged Nariai black hole is quickly screened by Schwinger pair production; it discharges instantaneously such that the black hole charge becomes $Q=0$ but its mass remains above the neutral Nariai mass limit $M^2\Lambda > 1/9$. This marks a super-extremal black hole solution, lying beyond the sharkfin, and thus in violation of the WCC. In the cosmologists' parlance, this is a ``Big Crunch" solution where the two-sphere collapses to zero. 

\par In contrast, if $m \gg qgM_PH$, the black hole gradually evaporates to empty dS space, as expected \cite{BoussoHawking1996_PairCreation,
Bousso1996_Nariai,
Bousso1999_QuantumStructredS,
MossNaritaka2021_KerrdSevaporation}. In Minkowski space-time, the prohibition of super-extremal black holes and the preservation of WCC leads to the WGC \cite{ArkaniHamed2006_WGCorigins}. If we follow these same principles in dS space-time, we must forbid $m \ll qgM_PH$. In so doing, Refs. \cite{vanRiet2019_FLevapBHdS,
vanRiet2021_FL} utilise the charged Nariai black hole as a laboratory in which to construct a dS analogue for the WGC, namely the FL bound of Eq. (\ref{eq:FLbound}).

\subsection{Scalar QNM potential with non-zero mass and charge in the RNdS space-time \label{subsec:potential}}
\par As a perturbed black hole settles towards an equilibrium state, it emits a discrete set of damped frequencies referred to as ``quasinormal frequencies" (QNFs)  \cite{Vishveshwara1970,Press1971}. These QNFs are independent of the initial perturbing stimulus and can be used to extract characteristic insights about their black hole source. In particular, the oscillation frequency $\mathbb{R}e \{ \omega \}$ and damping $\tau = -1/\mathbb{I}m \{\omega \}$ of the QNFs are uniquely determined by the black hole mass, charge, and spin \cite{Echeverria1989_BHpropertiesestimate}. With the LVK collaboration's rising success in extracting these QNFs from GW observations of compact binary coalescence, QNM studies can now be applied in the search for deviations in GR through novel testing of the no-hair conjecture \cite{LIGO2019_GWTC1-GRtest,LIGO2020_GWTC2-GRtest_pyRing3,LIGO2021_GWTC3-GRtest,Carullo2019_pyRing1,refNoHair_pyRing2}. Such investigations allow for an exploration of gravity in the relatively-untested strong regime, complementing extant results obtained from experiments in low-velocity linear regimes \cite{Stairs2003_GRpulsarTests,Will2009_GRstellarTests,refWillReview2014,Berti2015_TestingGRastro}. QNMs have far-reaching applications including quantum gravity conjectures; modified theories of gravity; stability analyses of naked singularities and novel space-times; the development of exact, semi-classical, and numerical tools for solving differential equations; the construction of numerical relativity simulations; the explorations of the gauge-gravity duality (see Refs. \cite{refNollert1999,refKokkotasRev,refFerrari2008,refBertiCardoso,
refKonoplyaZhidenkoReview,Grandclement2007_Spectral,Dias2015_Num} for further reading).

\par In this section, we shall focus on a mathematical construction of the QNM problem. We can approach the QNM calculation as an eigenvalue problem subjected to physically-motivated boundary conditions that follow from considering the space-time of interest through a classical lens. As stated in the introduction, radiation is considered to be purely ingoing at the event horizon and purely outgoing at the dS horizon for a dS space-time. On the basis of spherical symmetry and time independence, the QNM behaviour in static black hole space-times can be shown to reduce to a simple radial wave equation, as first demonstrated explicitly in Refs. \cite{refRW,refZerilli} for the Schwarzschild case. 

\par Let us consider a minimally-coupled scalar field with a non-zero mass and charge, oscillating on a RNdS background with a harmonic time dependence. The QNM can be written in terms of its temporal, radial, and angular components,
\begin{align} \label{eq:ansatz}
    \Psi_{n \ell m} (t,r,\theta,\phi) = \sum_{n=0}^{\infty}\sum_{\ell,m}^{\infty} \frac{\psi_{n\ell m} (r)}{r} \; Y_{\ell m} (\theta,\phi) \; e^{-i \omega_{n \ell m} t} \;.
\end{align}
Here, the angular contribution is expressed using spheroidal harmonics, for which $\ell$ and $m$ represent the angular momentum (multipolar) and azimuthal numbers. For each $\ell$, there exists infinitely many overtones $n \geq 0$ labelling the QNF in increasing multiples of $\mathbb{I}m \{ \omega \},$ with the $n=0$ as the ``fundamental mode" representing the least-damped and thus longest-lived QNM. 

\par Eq. (\ref{eq:ansatz}) satisfies the Klein-Gordon equation in curved space-time for a massive charged scalar test field,
\begin{equation} \label{eq:KG}
\frac{1}{\sqrt{-g}} \left( \partial_{\mu} - iq A_{\mu} \right)\left( \sqrt{-g} g^{\mu \nu} \left( \partial_{\nu} - iq A_{\nu} \right) \psi \right) = \mu^2 \psi \;,
\end{equation}
\noindent where $g$ refers to the metric defined in Eq. (\ref{eq:metric}), $\mu$ and $q$ represent the mass and charge, respectively, and $A_t(r)~=~-~Q/r~dt$ is the electrostatic four-potential of the black hole as defined in Eq. (\ref{eq:vecA}). Both $\mu = \mu c /\hbar$ and $q = qc/\hbar$ have units of inverse length in natural units. In these units, $q$ should be a multiple of the electron charge $\vert e \vert \sim$ 0.1. The radial component of the Klein-Gordon equation can then be written as
\begin{equation}
\frac{d}{dr}\left(r^2 f(r)\frac{dR}{dr}\right)+\left(\frac{r^2(\omega+qA_t(r))^2}{f(r)}-\ell(\ell+1)-\mu^{2}r^2 \right) R(r)=0\,. \label{eq:radial}
\end{equation}%
\noindent Then, redefining $R(r)$ as $R(r)=F(r)/r$ and employing the tortoise coordinate $r_*$,\footnote{The general \enquote{tortoise coordinate} is defined as
$d/dr_* = f(r) d/dr ,$
where $r_*=r_*(r)$ serves as a bijection from $(r_+,r_c)$ to $(-\infty,+\infty)$. If we set $r_*(r=0)=0$, the general tortoise coordinate for the RNdS black hole becomes
\begin{equation}
r_* (r)=\int \frac{dr}{f(r)} = \sum^{4}_{i=1} \frac{1}{2\kappa_i} \ln \left(1- \frac{r}{r_i} \right) \, ,
\end{equation}
\noindent where $\kappa_i$ is the usual surface gravity at a horizon \cite{refIKrn,refHorowitzChap1}.} we obtain
\begin{equation} \label{eq:ode}
\frac{d^{2}F(r_*)}{dr_*^{2}} + \left[ \omega^2 -V(r) \right] F(r_*)=0 \;,
 \end{equation}
\begin{equation}\label{eq:pot}
\mathrm{with} \;\;\; \; V(r)  = f(r) \left[ \frac{\ell (\ell + 1)}{r^2}  +   \frac{f^\prime(r)}{r} + \mu^2 \right] -2\omega q A_t(r)-q^2A_t(r)^2 \;.
 \end{equation}

\par
\begin{figure}[t!]
    \centering
        \includegraphics[width=0.7\linewidth]{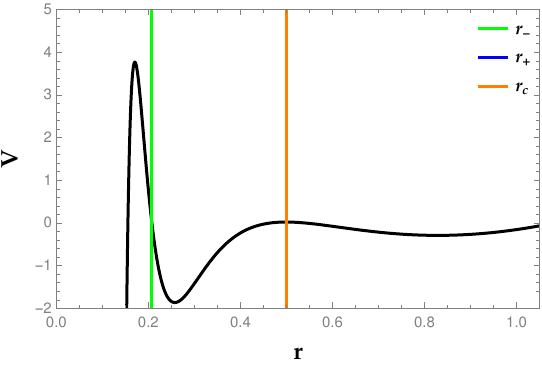}
        \caption{\textit{
        The QNM potential of a massive, charged scalar test field with $L^2_{dS} = 3 / \Lambda =1$, $\ell=1$, $q=0.1$, and $\mu =0.1$. Here, $M=Q=1/4$, corresponding to point $E$ on Fig. \ref{fig:sharkfin}: the charged Nariai solution where $M=Q=1/4$ and $r_+ = r_c$. The green line indicates the Cauchy horizon $r = r_c$; orange corresponds to the $r_+ = r_c$ line. For the ultracold case (point $U$), all horizons converge on this line at $r =3M/2 \sim 0.408$ by Eq. (\ref{eq:ultracold}).
}}
        \label{fig:RNdSpotential-chNariai}
\end{figure}

 In Fig. \ref{fig:RNdSpotential-chNariai}, we illustrate the behaviour of the potential with respect to the sharkfin of Fig. \ref{fig:sharkfin} for the charged Nariai case. QNMs are defined on the black hole exterior $r \in (r_+ , r_c)$, framed by the blue and orange delineators; the green line indicates the inner horizon $r=r_-$. We set $L^2_{dS} = 3 / \Lambda =1$, $\ell=1$, $q=0.1$, and $\mu =0.1$.  Note that in Fig. \ref{fig:RNdSpotential-chNariai}, there is no physical exterior black hole region. QNM analyses are therefore impossible. In this configuration, we see explicitly the effect of the cosmological constant on the parameter space: the $M=Q$ solution, which in asymptotically-flat space corresponds to a coalescence of the Cauchy and event horizons, is associated with a meeting of the event and dS horizons and only for the maximal value of $(M=Q)_{max}$. At this point, the dS space-time is dominated entirely by the black hole.
\par With the \href{https://github.com/anna-chrys/RNdS_QNMs}{interactive Mathematica notebook}, the QNM potential corresponding to other regions of the phase space can be explored. For a fixed $Q$, the amplitude of the potential decreases significantly as we increase $M$ from the near-extremal $OU$ branch to the near-Nariai $NU$ branch. Despite this, the potential retains its shape. This seems to suggest that QNM computational methods dependent on barrier potentials (e.g. WKB-based, potential-based, and photon-orbit techniques) may be applicable throughout the sharkin. We can also explore the effect of the scalar field parameters on the barrier potential. For example, raising $\ell$ increases the magnitude of $V(r)$; raising $q$ elevates the potential slightly but varying $\mu$ negligibly affects the region of interest $r_+ < r < r_c$. $V(r_{peak})$ is largest near point $O$ and decreases towards $U$, remaining largest along the extremal branch $OU$.

\par Finally, we consider how the potential changes as we cross the $M=Q$ line. For warmer $M>Q$ solutions, the event and dS horizons are relatively close and the potential's amplitude is suppressed; for colder $M<Q$, the parameter space beyond the event horizon as well as the amplitude of the potential increase.  

\begin{figure}[t]
    \centering
        \includegraphics[width=0.7\linewidth]{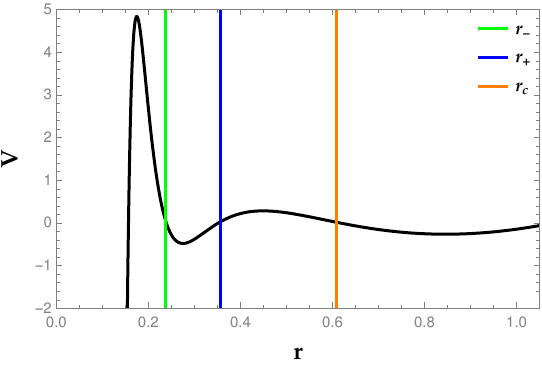}
    \caption{\textit{
    The QNM potential of a massive, charged scalar test field with $L^2_{dS} = 3 / \Lambda =1$, $\ell=1$, $q=0.1$, and $\mu =0.1$. Here, $M=0.2425$ and $Q=0.249$ such that $Q>M$, corresponding to the ``cold" black hole case 
    Green, blue, and orange delineators indicate Cauchy, event, and cosmological horizons, respectively.
    }}
           \label{fig:RNdSpotential-cold}
\end{figure}

\par For a massive charged scalar field in a RNdS background, the QNM boundary conditions can then be expressed concisely \cite{KonoplyaZhidenko2014_RNdSrprc,Hod2018_RNdS,
DiasReallSantos2018_SCC,DiasSantos2020_RNdSinstability} as

\begin{equation} \label{eq:BCdS}
\psi \sim 
\begin{cases}
e^{-i \left(\omega - \frac{qQ}{r_+} \right)r_*} \;, & \quad r \rightarrow r_+ \;\; (r_* \rightarrow - \infty) \;,\\
e^{+i\left(\omega - \frac{qQ}{r_c} \right)r_*} \;, & \quad r \rightarrow r_c \;\; (r_* \rightarrow + \infty) \;.
\end{cases}
\end{equation} 
\noindent These QNM boundary conditions are applied to perturbations exterior to the event horizon. Recall that, classically, the event horizon serves as a one-sided membrane through which energy is lost to the black hole interior. Furthermore, energy cannot enter into the system from beyond the dS horizon. Consequentially, the system is not time-symmetric; the eigenvalue problem is non-Hermitian and the eigenvalues are complex. The corresponding eigenfunctions are then not normalisable and do not form a complete set (see reviews \cite{refNollert1999,refBertiCardoso,refKonoplyaZhidenkoReview} for further discussion).

\par To confront these technical difficulties, a host of techniques have been developed to calculate QNFs, including ``exact" (e.g. direct integration methods \cite{DavisPrice1971,Price1994}, the continued fraction method \cite{refLeaver1985}, inverse-potential methods that approximate the effective potential with an inverse P{\"o}schl-Teller potential \cite{PoschlTellerPotential} for which bound-state solutions are known \cite{PoschlTellerMethod}, pseudospectral methods \cite{Dias2009_Paraspectral1,Dias2009_Paraspectral2}, etc.) and numerical (e.g. the asymptotic iteration method \cite{refAIM_OG,refAIM}, the Horowitz-Hubeny technique \cite{HorowitzHubeny}, etc.) methods. There are also several semi-classical tools available (e.g. WKB-based methods ideal for calculating QNFs in the $\ell \geq n$ regime \cite{refBHWKB0,refBHWKB0.5,refBHWKB1} at sixth-order \cite{Konoplya2003} and beyond \cite{Konoplya2019}, photon-orbit methods such as the inverse multipolar expansion method \cite{refDolanOttewill2009} that exploits the known relationship between QNMs and unstable null geodesics \cite{refGoebel1972} to build an ansatz with which the QNM problem can be solved iteratively with increased accuracy for large values of $\ell$ \cite{refOurLargeL}, etc.). This is hardly an exhaustive list; see Refs. \cite{refBertiCardoso,refKonoplyaZhidenkoReview,Grandclement2007_Spectral,Dias2015_Num} for further details.
   
\par In this treatment of QNMs as an eigenvalue problem, we refer to the QNMs themselves as the discrete set of eigenfunctions which satisfy the above boundary conditions; the corresponding eigenfrequencies are the countable infinity of complex QNFs. As already mentioned, these can be decomposed into a real and an imaginary part where the former represents the physical oscillation frequency and the latter is proportional to the (inverse) damping time of the QNM. The imaginary part of the QNF must be negative to ensure black hole stability \cite{Vishveshwara1970stable}. We shall return to this point in section \ref{subsec:superrad}.

\section{Scalar QNFs with non-zero mass and charge \label{sec:QNFs}}

\par As first stipulated in Ref. \cite{Cardoso2017_QNMsSCC}, and further explored in Refs. \cite{DiasReallSantos2018_SCC,
Mo2018_SCC,
DiasSantos2020_RNdSinstability,
Papantonopoulos2022}, electrically-neutral scalar QNFs can be classified into three qualitatively distinct families based on the structure of their QNF solution: a ``photon-sphere" family, a ``dS" family, and a ``near-extremal" family. The photon-sphere (dS) family connects smoothly to the QNMs of asymptotically-flat (empty dS) space-time; the near-extremal family, on the other hand, is unique to the RNdS space-time. An additional ``orphan mode" was also identified for $\ell = 0$, which we discuss further in section \ref{subsec:superrad}. We find that each of these can be loosely associated with a particular region in the sharkfin. With this in mind, we can categorise the QNFs according to their structure \cite{Cardoso2017_QNMsSCC} and position in the phase space, $viz$
\begin{outline}
\1[$(i)$] near-extremal modes\footnote{In discussions on the SCC, near-extremal modes are the most relevant. Since the charged Nariai branch is the focus of discussions on the FL bound and the WGC, we are interested primarily in the photon-sphere modes in this work.} arise near the $OU$ line where $r_- \sim r_+$, 
\begin{equation}
\omega_{NE} \approx - i (\ell + n + 1) \kappa_- = - i (\ell + n + 1) \kappa_+ \;;
\end{equation}
\1[$(ii)$] dS modes can be found near Point $O$, following closely along branch $OU$ and in competition with the near-extremal modes. Here, $\kappa_c \sim 1/L_{dS}$ and
\begin{equation}
\omega_{dS_{n=0}} \approx -i \ell  \kappa_c \;, \quad \omega_{dS_{n \neq 0}} \approx -i (\ell + n + 1) \kappa_c \;;
\end{equation}
\1[$(iii)$] photon-sphere modes occupy a wide space beneath the $M=Q$ line and approach the $NU$ line, characterised by large $\mathbb{R}e \{ \omega \}$ and for which $\mathbb{I}m \{ \omega \}$ is related to the instability time scale of null geodesics near the black hole photon sphere in the eikonal regime. Closely following the charged Nariai $NU$ branch for smaller $Q$ values,
\begin{align}
    \mathbb{I}m \{ \omega_{PS} \} \approx -i \left(n + \frac{1}{2} \right) \kappa_+ \;.
\end{align}
\end{outline} 
\noindent We note that we observe a non-zero $\mathbb{R}e \{ \omega \}$ part in each of these regions. However, in the case of the near-Nariai region, this contribution is $\mathbb{R}e \{ \omega \} \sim \mathcal{O}(0.01)$ for $Q<0.1$.
\par In Ref. \cite{Cardoso2017_QNMsSCC}, for $q=0$, a study of the photon-sphere modes revealed two sets of QNFs for which the imaginary parts were identical and the real parts were of equal magnitude but opposite sign. This is a natural consequence of the QNM reflection symmetry $\omega \rightarrow - \omega^{\star}$, such that the complex conjugate $\omega^*$ is the QNF corresponding to QNM $\psi^*$ satisfying Eq. (\ref{eq:ode}).

\par However, this degeneracy is broken when $q \neq 0$ is introduced. This was observed in the semi-classical analysis of Ref. \cite{DiasReallSantos2018_SCC}, and in other works e.g. Refs. \cite{Mo2018_SCC,Papantonopoulos2022}. In the computation in Ref. \cite{DiasReallSantos2018_SCC} of the QNF spectrum in the large-$q$ regime, the authors found that their semi-classical method predicted two ``families" of solutions: a black-hole family $\omega_+$ and a cosmological-horizon family $\omega_c$.\footnote{Dias, Reall, and Santos' analysis assumed a large $q r_c$ with respect to $\mu r_c$ and $\ell$, and employed a technique based on the inverse-multipolar expansion method of Dolan and Ottewill, first conceptualised in Ref. \cite{refDolanOttewill2009}. The Kerr analysis of Dolan in Ref. \cite{Dolan2010_KerrQNMs} employs this same method, albeit with an additional series expansion in the angular eigenvalue. Inspired perhaps by the analogous relationship between RN and Kerr black holes, Dias \textit{et al.} used an expansion in inverse powers of $q$ for their QNF analysis based on the computation in Ref. \cite{Dolan2010_KerrQNMs}.} The physical meaning of these two families, however, is not addressed.  In this work, we focus instead on the small-$q$ regime, and employ the semi-classical method of Ref. \cite{Papantonopoulos2022}.\footnote{In particular, the small $qM$, small $Q/M$, and large-$\ell$ regime.}   
 
\subsection{QNFs in the large-$\ell$, small-$Q$ regime: a semi-classical treatment}

\par The application of semi-classical methods to QNF calculations has a long and successful history, dating back to the WKB analyses of Schutz, Iyer, and Will \cite{refBHWKB0,refBHWKB0.5,refBHWKB1} and receiving significant improvements since (see Ref.  \cite{Konoplya2019} and references therein). Building upon these, and the well-cultivated relationship between the bound states of anharmonic oscillators \cite{BenderWu1969_AnharmonicOscillator} and black hole QNFs \cite{PoschlTellerMethod,refFerrMashh1,refFerrMashh2},  Hatsuda proposed a Borel-resummation method that allows for a simple means of calculating the QNF spectrum in spherically-symmetric black hole space-times \cite{Hatsuda2019_WKB}. This technique was applied to the RNdS black hole space-time by Gonz{\'a}lez \textit{et al.}, who computed the QNFs in the eikonal regime for small $Q/M$ and $qM$ \cite{Papantonopoulos2022}. To do so,  the QNF can be written as a series expansion in inverse powers of $L = \sqrt{ \ell (\ell + 1)}$,
\begin{equation} \label{eq:omegaseries}
\omega = \sum_{k = -1} \omega_k L^{-k} \;.
\end{equation} 
\noindent The series expansion is then inserted into
\begin{equation} \label{eq:BorelQNF}
\omega =\sqrt{V(r^{max}_*)-2 i U} \;, \quad
U \equiv U(V^{(2)},V^{(3)},V^{(4)},V^{(5)},V^{(6)})  \;,
\end{equation}
\noindent where $U$ is given explicitly in Eq. (A-4) of Ref. \cite{Papantonopoulos2022} (see appendix \ref{app:num} for details). The objective of the method is to solve for the $\omega_k$ coefficients for increasing orders of $k$. $V(r^{max}_*)$ corresponds to the peak of the barrier potential, located at 
\begin{equation}
 r^{max}_* \approx r_0 + r_1 L^{-2} + ...\;, \quad \text{with} \quad V(r^{max}_*) \approx V_0 + V_1 L^{-2} + ... \;, 
\end{equation}
where subscripts refer to terms in the series expansion. The numbered superscripts in Eq. (\ref{eq:BorelQNF}) refer to derivatives $V^j$, taken with respect to a generalised tortoise coordinate, such that
\begin{equation}
V^{j} = \frac{d^j V(r^{max}_*)}{dr^j} = f(r) \frac{d}{dr} \left[ f(r) \frac{d}{dr} \left[... \left[f(r) \frac{d V(r)}{dr} \right]... \right] \right]_{r \rightarrow r^{max}_*} \;.
\end{equation}

\par As already stipulated, this method is most reliable in the large-$\ell$ regime, and for small values of $Q$ and $q$. Beyond these limitations, the method allows us to maintain the black hole and scalar field parameters as free variables. In other words, the iterative procedure required to produce an expression for the QNF, Eq. (\ref{eq:omegaseries}), can be computed with reasonable accuracy without pre-defining the masses, charges, overtones, or harmonics (which is required for many other semi-classical methods e.g. consider Refs. \cite{Konoplya2019,KonoplyaZhidenko2022_Bernstein} and references therein). As a result, we can observe the evolution of the QNFs as we move throughout the RNdS phase space via the \href{https://github.com/anna-chrys/RNdS_QNMs}{interactive Mathematica notebook}.

\par An interesting effect noted in Fig. 3 of Ref. \cite{Lagos2020_Anomalous} for massive scalar modes in the Schwarzschild (and Kerr) black hole space-time is a ``critical mass" value: a common point of intersection irrespective of the hierarchy in $\ell$ in the $-Im (\omega)$ vs $\mu$ graph. This is observed also for the charged, massive scalar field in the RNdS space-time: in Figs. 9 and 10 of Ref. \cite{Papantonopoulos2022}, where the lines intersect of low values of $q$. A shared point of intersection corresponds to a critical mass $\mu_{crit}$, which can be solved for by setting $\mathbb{I}m \{ \omega_{-2} \} =0$ \cite{Papantonopoulos2022} (see appendix \ref{app:num}). At the lowest order in $Q$, 
\begin{equation} \label{eq:mucritLO}
   \mu_{crit}^2 =  \frac{18045 \Lambda  M^2+137}{29160 M^2} \;.
\end{equation}

\begin{figure}[t]
\centering
\includegraphics[width=0.8\linewidth]{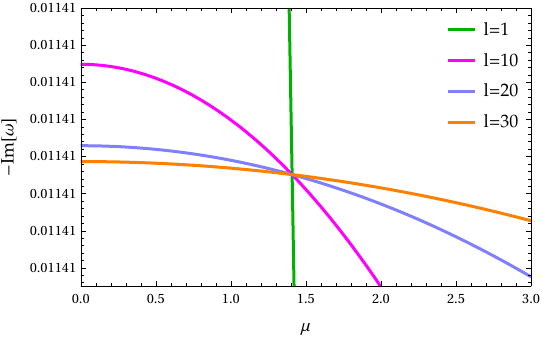}
\caption{\textit{ A graphical indication of the critical mass $\mu_{cit}$ at which $\mathbb{I}m \{ \omega \}$ values converge irrespective of $\ell \geq 1$. At Point $N$, where $r_+ \sim r_c$ and $(M,Q)=(1/\sqrt{27},0)$, $\mu_{crit} \sim \pm 1.4083$ for very small $\mathbb{I}m \{ \omega \} \sim -0.0114$, and $M \mu \sim 0.3$; this is the smallest possible $\mu_{crit}$ for the RNdS black hole.}}
\label{fig:ImwVSmuNariai}
\end{figure}

\begin{figure}[t]
    \centering
    \begin{subfigure}[t]{0.45\textwidth}
        \centering
        \includegraphics[width=\linewidth]{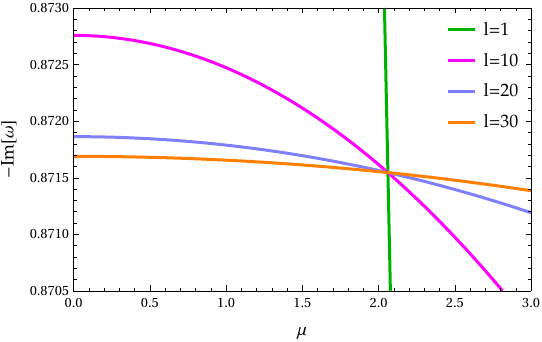}
        \caption{\textit{ For $M=Q=0.104$ and $q=0$, the critical mass is $\mu_{crit} \sim \pm 2.065$.}}
\label{fig:ImwVSmu_MQ}
    \end{subfigure}%
    ~ \hspace{0.5cm}
    \begin{subfigure}[t]{0.45\textwidth}
        \centering
        \includegraphics[width=\linewidth]{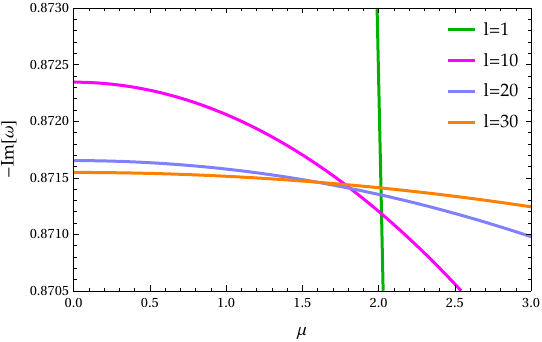}
        \caption{\textit{ For $M=Q=0.104$ and $q=0.1$ there is no longer an intersection point.}}
    \label{fig:ImwVSmu_MQc01}
    \end{subfigure}
    \caption{\textit{Increasing the charge from $q=0$ to $q=0.1$ displaces the intersection. Our results suggest that a critical $\mu_{crit}$ can only be found in the $\mu M \gg qQ$ regime.}}
    \label{fig:critmass}
\end{figure}

\par As pointed out in Refs. \cite{Lagos2020_Anomalous,Papantonopoulos2022}, the critical mass value is associated with an unusual damping behaviour. It is expected that an increase in $\ell$ or an increase in $\mu$ leads to increased oscillation frequency and decreased damping, such that the magnitude of the real part increases and the magnitude of the imaginary part decreases \cite{refBertiCardoso,refKonoplyaZhidenkoReview}. For $\mu < \mu_{crit}$, this is observed. However, beyond $\mu \geq \mu_{crit},$ the magnitude of the imaginary part increases with larger values of $\ell$. This can be seen explicitly in Fig. \ref{fig:ImwVSmuNariai}, where $\vert \mathbb{I}m \{\omega_{\ell=10} \} \vert > \vert \mathbb{I}m \{\omega_{\ell=30} \} \vert$ for $\mu < \mu_{crit}$ and $\vert \mathbb{I}m \{\omega_{\ell=10} \} \vert < \vert \mathbb{I}m \{\omega_{\ell=30} \} \vert$ for $\mu > \mu_{crit}$. We observe that this coalescence indicates a negligible dependence on the $L$ for critical mass $\mu_{crit}$. We can explore graphically for which values of $q$, $Q$, and $M$ we obtain this intersection using the \href{https://github.com/anna-chrys/RNdS_QNMs}{interactive Mathematica notebook}. For example, we demonstrate in Fig. \ref{fig:ImwVSmuNariai} that for the uncharged Nariai case, the critical mass corresponds to $\mu_{crit} \sim 1.4083$ for all $q$. This same result is found using Eq. (\ref{eq:mucrit}) for $q=0$. This is the maximum $M$ for which we see an intersection. We do not observe further intersections along the $NU$ branch, and since $Q=0$, the charge $q$ has no influence on the QNF at point $N$. For $q \approx 0$, the intersection point shifts right as we increase $Q$, and as we increase $M$ (from zero) the intersection point shifts left. For fixed values of $M$ and $Q$, the intersection point shifts left as we increase $q$ (along the small domain of $q$). 

\par When $q=0$, the influence of $Q$ is negligible on $\mu_{crit}.$ We find that large $M$ corresponds to a small $\mu_{crit}$; $\mu_{crit}$ decreases with increasing $M$, such that for $M \sim 0$, $\mu_{crit} \sim 20$. When $M > 1/\sqrt{27 \Lambda^2}$ for non-zero $Q$, we observe growing rather than decaying modes for large values of $\mu$ e.g. when $\mu > 5$ for $\ell \sim 10$ and $\mu > 15$ for $\ell - 30$. For $q>0,$ we do not observe an intersection but the anomalous behaviour in which $\vert \mathbb{I}m \{\omega_{\ell} \} \vert > \vert \mathbb{I}m \{\omega_{\ell+1} \} \vert$ is noted for $\mu > 0.$ As we would expect from the coupling between $q$ and $Q$, $Q$ has a more obvious effect for the nonzero $q$, such that the spacing between modes increases with $Q$.   

\subsection{Instability and superradiance \label{subsec:superrad}}

\par A black hole that is unstable against small perturbations will be short-lived: it is destined to disappear or transform into some other object. Stability is thus a prerequisite for its continued existence and, by extension, an essential assumption for any examination of phenomena in the black hole space-time. In the case of higher-dimensional space-times, which do not benefit from a black hole uniqueness theorem and therefore admit a variety of topologies \cite{refHorowitzBook}, stability analyses serve as this criterion for existence. 

\par We can determine if a black hole is stable against the perturbations of a particular test field through an analysis of the QNF spectrum. Assuming a harmonic time dependence $\psi \sim e^{-i \omega t}$, a negative imaginary component reveals an exponentially decaying system, in accordance with a return to an equilibrium state \cite{Vishveshwara1970stable}. An unstable QNM can be identified by a positive imaginary component, indicative of an exponential growth in the oscillations. For the RNdS black hole, stability had been established for $d < 7$ for massless fields \cite{refIKrn,refIKchap6,refKonoplya2009}. However, as explained in section \ref{sec:intro}, attention was refocused on the stability of the 4D RNdS black hole in the wake of the observed instability for the $\mu = \ell = 0$ case \cite{ZhuZhang2014_RNdSinstability}. A further RNdS instability was observed in Ref. \cite{DiasReallSantos2018_SCC} for a charged scalar field when $\mu = 0$ and $q r_c \ll 1$.

\par An exponentially growing mode can be caused by superradiance. In the black hole context, classical superradiance was first addressed in the case of a rotating (Kerr) black hole \cite{Starobinsky1973_SR1,Starobinsky1974_SR2}, where rotational energy is carried away from the spinning black hole and enhanced. This becomes especially interesting for bosonic fields with a non-zero mass $m=\mu \hbar$. In this case, the superradiant amplification can be augmented by the confining potential of the massive bosonic field, leading to a runaway self-amplification of the field and the formation of a bosonic condensate around the black hole source. Though this process has been studied for several decades (see Refs. \cite{refBertiCardoso,refKonoplyaZhidenkoReview,BritoCardosoPani2020_Superradiance}), it has become a subject of renewed interest due to the possibility of detecting such a phenomenon through GW astronomy \cite{BertiBrito2017a_UltraLightStochastic}. Since the bosonic cloud and its black hole source both encode information about the perturbing field, GW observations could in principle be used to constrain the mass of the bosonic field. This is an active area of research, where GW analyses performed in Refs. \cite{BertiBrito2017b_UltraLightLIGOLISA,BertiBrito2019_UltraLight1GW} suggest that LIGO and LISA may be able to probe the parameter space of light bosons at $m \sim 10^{-13}-10^{-11}$ eV and $m \sim 10^{-19}-10^{-14}$ eV, respectively. These constraints overlap well with hypothetical ``string axiverse" particles \cite{Arvanitaki2009_StringAxiverse,PDG2022}.  

\par When the Compton wavelength of the massive bosonic field is of the order of the black hole’s radius, the scaling for the suppression of the instability timescale is governed by the dimensionless parameter $M \mu \lesssim 1$ (recall that $M$ is the length scale for the black hole and $\mu$ has units of inverse-length). In the case of ultralight scalar fields, this instability timescale is of the order of seconds (hundreds of years) for black holes of mass $M \sim M_{\odot}$ $(M \sim 10^9 M_{\odot})$  \cite{refKonoplyaZhidenkoReview,BritoCardosoPani2020_Superradiance}. As discussed in Ref. \cite{Chrysostomou2022_BHnilmanifold}, reintroducing the SI units such that $M = m_{\rm BH} G /c^2$ and $\mu = mc/\hbar$ allows us to express the mass of a bosonic field as
\begin{equation}
    m = \frac{1}{m_{\rm BH}} \frac{\hbar c}{G} M\mu \;.
\end{equation}
\noindent Since $M \mu \sim \mathcal{O}(1),$ $\hbar c /G \sim 10^{-16} $ kg$^2$, and $1 M_{\odot} \sim 10^{30}$ kg, we can scale the black hole mass as $m_{\rm BH} \sim 10^\chi M_{\odot}$. Returning to natural units with $c=1$ (c.f. Eq. (4.25) of Ref. \cite{BritoCardosoPani2020_Superradiance}), we have
\begin{equation} \label{eq:Mm}
m \sim 10^{-( \chi + 10)}M \mu \text{ eV} \;.
\end{equation}
\noindent In the previous section, we have seen that for the critical mass in the regime $\mu M \gg q Q$, $\mu_{crit} M \leq 0.3$.  To satisfy the FL bound of $m > 10^{-3}$ eV, we can only consider compact objects corresponding to $\chi < -8$, such as micro black holes. This suggests that the FL bound rules out the possibility of observing weakly-charged scalar QNMs from astrophysical black holes.

\par
\begin{figure}[t!]
    \centering
        \includegraphics[width=0.7\textwidth]{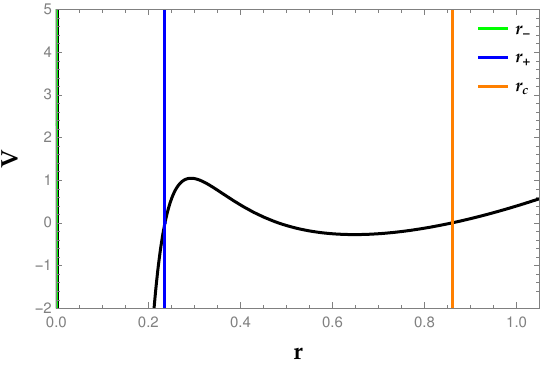}
     \caption{\textit{The QNM potential of a massive, charged scalar test field with $L^2_{dS} = 3 / \Lambda =1$, $\ell=0$, $q=0.1$, and $\mu =0.1$. We set $M=0.112$ and $Q=0.016$ such that the potential corresponds to a position in the phase space central to $ON$. For fixed $Q$, increasing $M$ shrinks the amplitude of the potential. Note the valley following the peak of the barrier potential, which allows for superradiant amplification of the reflected wave. For $\ell >0$, these disappear, as seen in Figs. \ref{fig:RNdSpotential-chNariai} and \ref{fig:RNdSpotential-cold}.  \label{fig:zeroL}} }
\end{figure}

 To a lesser extent, superradiance has also been studied in the context of charged black holes. The connection between the rotating and charged black holes lies in the so-called ``charge-angular momentum analogy", where near-extremal charge corresponds to fast rotating black holes \cite{Cardoso2017_QNMsSCC}. Superradiant instability for charged fields within RNdS space-times can be inferred directly from a study of the QNM effective potential: as we see in Fig. \ref{fig:zeroL}, a local minimum follows the local maximum and both appear within the physically-relevant domain $r_+ < r < r_c.$ A wave scattering off the potential barrier will become amplified in this ``valley", resulting in a superradiance that destabilises the black hole.

\par We can explore this further, following Refs. \cite{KonoplyaZhidenko2014_RNdSrprc,
ZhuZhang2014_RNdSinstability}. Consider the scattering (rather than purely QNM) problem: the ingoing wave from $r=r_c$ partially passes through the potential barrier, passing $r = r_+$ to fall inside the event horizon, while the rest is reflected from the potential barrier back towards the cosmological horizon $r= r_c$. The boundary conditions of Eq. (\ref{eq:BCdS}) are modified
, such that
\begin{equation} \label{eq:BCdSrt}
\psi \sim 
\begin{cases}
T e^{-i \left(\omega - \frac{qQ}{r_+} \right)r_*} \;, & \quad r \rightarrow r_+ \;\; (r_* \rightarrow - \infty) \;,\\
e^{-i \left(\omega - \frac{qQ}{r_c} \right)r_*} + R e^{+i\left(\omega - \frac{qQ}{r_c} \right)r_*} \;, & \quad r \rightarrow r_c \;\; (r_* \rightarrow + \infty) \;.
\end{cases}
\end{equation} 
\noindent $R$ is the amplitude of the reflected wave (i.e. reflection coefficient), and $T$ is the transmitted-wave amplitude (i.e. transmission coefficient). From the Wronskian (constant, implying linear independent solutions),
\begin{equation}
1-|R|^2 = \frac{\omega - qQ/r_+}{\omega - qQ/r_c} |T|^2 \;.
\end{equation} 
\noindent Note the coefficient of $|T|^2$. 

\par This relationship between superradiance and instability for charged fields can be presented concisely following Ref. \cite{KonoplyaZhidenko2014_RNdSrprc}. To begin, we multiply the generic ordinary differential equation, Eq. (\ref{eq:ode}), by its complex-conjugate $F^*$, and apply integration-by-parts
\begin{align}
    F^*(r_*) \frac{dF}{dr_*} \Bigg \vert^{+\infty}_{-\infty} &+ \int^{+\infty}_{-\infty} (\omega + qA_t(r))^2 \vert F(r_*) \vert^2 dr_* \nonumber \\
    & =  \int^{+ \infty}_{- \infty} \left( V(r) \vert F(r_*) \vert^2 + \bigg \vert \frac{dF}{dr_*} \bigg \vert^2 \right) dr_* \;.
\end{align}
\noindent Since the potential is positive, the right-hand-side of the above expression is positive. Moreover, $\vert qA_t (r) \vert$ is monotonically increasing. Suppose we take only the imaginary parts of both sides, then if 
\[ \mathbb{R} e \{ \omega \} \geq  \frac{qQ}{r_+}\geq \frac{qQ}{r}  \quad \text{or} \quad \mathbb{R} e \{ \omega \} \leq  \frac{qQ}{r_c}  \leq \frac{qQ}{r} \;, \]
\noindent then $\mathbb{I}m \{ \omega \} < 0$ and the space-time is stable. From this, we obtain the superradiant condition,
\begin{equation} \label{eq:superradiance}
\frac{qQ}{r_c} < \mathbb{R}e \{ \omega \} < \frac{qQ}{r_+} \quad \implies \mathbb{I}m \{ \omega \} > 0 \;.
\end{equation}

\par Growing modes meet the Eq. (\ref{eq:superradiance}) criterion more easily, which reinforces the idea that superradiance is the mechanism behind this instability. For small $\Lambda$, a very small non-zero $\mu$ is sufficient to stabilise the field \cite{DiasReallSantos2018_SCC}, and a larger $\Lambda$ in turn requires a larger $\mu$. Here, we find that it is necessary to have a non-negligible contribution from $q$ in order to satisfy Eq. (\ref{eq:superradiance}), where a non-zero $\mu > 0$ contributes to increasing the QNF. We find in particular that in the shaded $OEU$ region, Eq. (\ref{eq:superradiance}) is satisfied for a wide range of $q$ and $\mu$ values. To justify this intuitively, it is worth remembering that this instability implies $qQ \gg \mu M$: the electromagnetic repulsion between particle and black hole exceeds their gravitational attraction \cite{KonoplyaZhidenko2014_RNdSrprc}.

\par
\begin{figure}[ht]
\centering
        \includegraphics[width=0.65\textwidth]{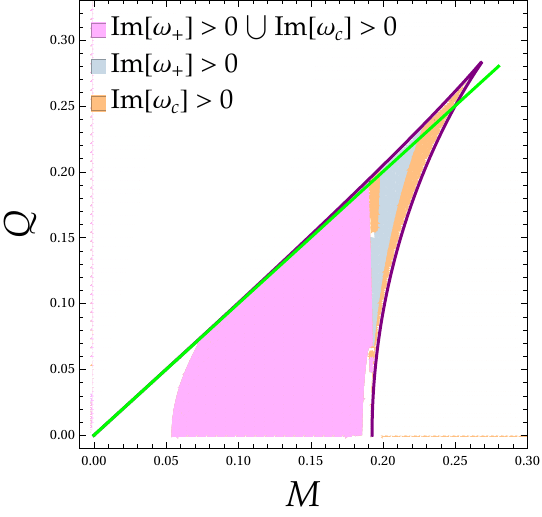}
        \caption{\textit{Regions of instability for $\omega_+$ (blue), $\omega_c$ (orange), and both (magenta), where we have set $q=0.1$, $\mu=10,$ and $\ell =1$. We observe that unstable regions emerge when the orders of magnitude of $\mu$ and $\ell$ differ.}  \label{fig:mu10c01l1unstable}}
\end{figure}

\begin{figure}[ht]
\begin{subfigure}[t]{0.45\textwidth}
        \centering
        \includegraphics[width=\textwidth]{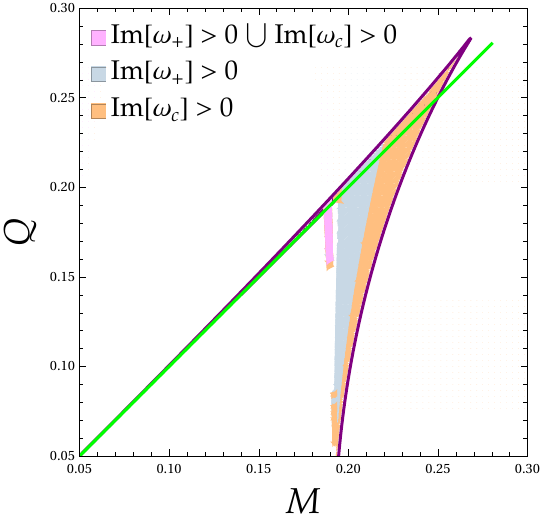}
        \caption{\textit{Regions of instability.}  \label{fig:mu10c01l10unstable}}
    \end{subfigure}%
    ~ $\quad$
    \begin{subfigure}[t]{0.45\textwidth}
        \centering
        \includegraphics[width=\textwidth]{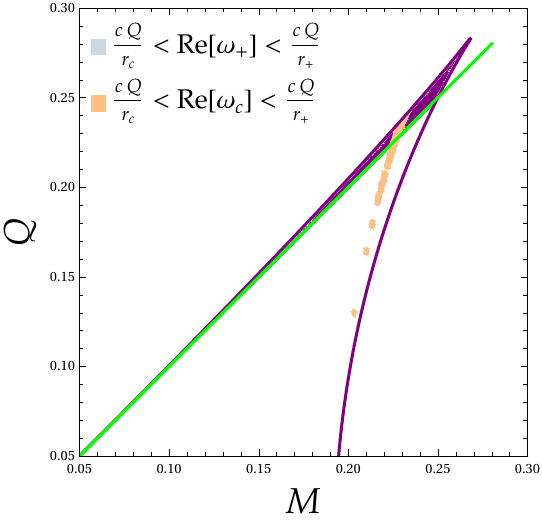}
        \caption{\textit{Regions reflecting superradiance}. \label{fig:mu10c01l10SR}}
    \end{subfigure} 
    \caption{\textit{We set the scalar field parameters as $q=0.1$, $\mu=10,$ and $\ell =10$ such that $\mu M \gg q Q$ and the QNFs are in the eikonal regime. Then we shade the regions within the RNdS parameter space for $L_{dS}=1$ whose QNFs indicate instability (Fig. \ref{fig:mu10c01l10unstable}) and superradiance (Fig. \ref{fig:mu10c01l10SR}). We distinguish between $\omega_+$ (blue), $\omega_c$ (orange), and both (magenta).}}
\end{figure}

 To satisfy $M \mu \sim 1$ for the $0 < M \leq \sqrt{2}/\sqrt{27}$ black hole parameter space, we set $\mu = 10$. When $q=0.1$  and $\ell = 1$, much of the sharkfin is shaded in favour of instability (Fig. \ref{fig:mu10c01l1unstable}) but there is no indication of superradiant modes within the sharkfin. If we set $\ell=10$ but maintain $\mu=10$ and $q=0.1$, only a small region $(M \sim 0.2$, $0.19 \lesssim Q \lesssim 0.21$, and only for $\omega_c)$ reflects superradiance (Fig. \ref{fig:mu10c01l10SR}). This, however, does correspond to the unstable region indicated in Fig. \ref{fig:mu10c01l10unstable} for $\omega_c$. Once we increase $q$, though, we observe evidence of superradiance. This is to be expected, as superradiance requires the coupling of the black hole and scalar field charge to overcome the coupling between black hole and field mass. Furthermore, we see in comparing Figs. \ref{fig:mu10c01l1unstable} and \ref{fig:mu10c01l10unstable} that increasing $\mu$ without increasing the angular momentum corresponding to the field results in a largely unstable QNF spectrum. Rather than evidence of a deeper relationship between $\mu$ and $\ell$, we concede that it may be simply that the large instability region is a numerical artefact. 

\subsection{On the issue of Strong Cosmic Censorship \label{subsec:SCC}}

\par As discussed in section \ref{sec:intro}, there is an expectation based on Penrose's SCC conjecture that the presence of a Cauchy horizon in the RN back hole interior leads to the infinite amplification of perturbations in the interior region through a blue-shift mechanism. The Cauchy horizon then is destabilised and behaves as a singularity beyond which the initial data cannot be extended. However, this is in direct conflict with the stability proven to hold for the RN black hole space-time for $d=4$. For the RNdS case, the presence of the cosmological horizon complicates matters, as the resultant ``red-shift" effect competes against the blue-shifting. This leads to a delicate balance between the damping of the perturbations in the exterior and the amplifications from the black hole interior. 

\par The condition that must be satisfied to prevent the red-shift from overcoming blue-shift, and thus preserve SCC, is provided in Eq. (\ref{eq:SCCvalid}), $viz.$
\begin{equation}
    -\frac{\mathbb{I}m \{ \omega^{n=0} \}}{\vert \kappa_- \vert} < \frac{1}{2} \;,
\end{equation}
\noindent where we would ordinarily expect that $\mathbb{I}m \{ \omega^{n=0} \}<0$, since this is a necessary condition for the stability of the black hole. It was found in Ref. \cite{DiasReallSantos2018_SCC} that for massless, uncharged scalar fields in RNdS space-times, $1/2 < \beta < 1$ for certain values of $M$ and $Q$, thereby indicating a SCC violation. 

\par We find that SCC is violated within the shaded $OEU$ region of the sharkfin and on certain points on the $OU$ line, particularly near $M \sim Q \sim 0.089 $ for near-zero mass and charge. Increasing $\mu$ decreases the magnitude of $\mathbb{I}m \{ \omega_+^{n=0} \}$ and $\mathbb{I}m \{ \omega^{n=0}_c \}$, and therefore could rescue SCC for a small enough value of $\beta$. On the other hand, increasing $q$ increases (decreases) the magnitude of $\mathbb{I}m \{ \omega^{n=0}_c \}$ ($\mathbb{I}m \{ \omega^{n=0}_+ \}$).   

\begin{figure}[ht]
\begin{subfigure}[t]{0.5\linewidth}
        \centering
        \includegraphics[width=\linewidth]{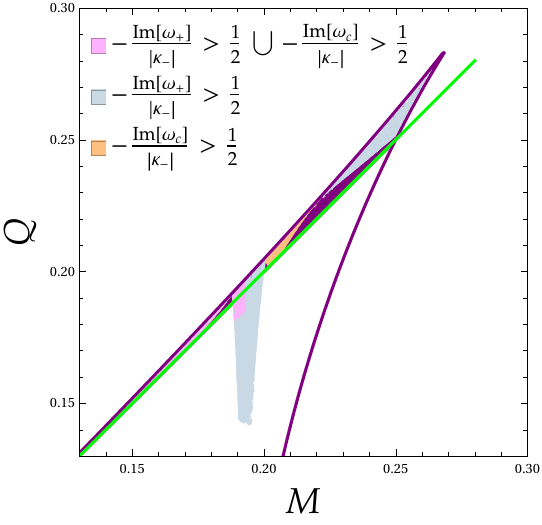}
        \caption{Regions where $\beta > \frac{1}{2}$ for $\mu=1$. \label{fig:SCCmu1c01l1}}
    \end{subfigure}%
    ~ 
    \begin{subfigure}[t]{0.5\linewidth}
        \centering
        \includegraphics[width=\linewidth]{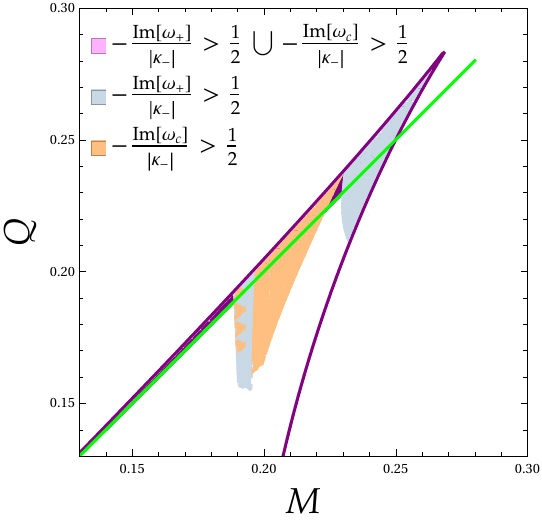}
        \caption{Regions where $\beta > \frac{1}{2}$ for $\mu=10$. \label{fig:SCCmu10c0}}
    \end{subfigure}
    \caption{\textit{We set the scalar field parameters as $q=0.1$ and $\ell =1$ such that $\mu M \gg q Q$. Then we shade the regions within the RNdS parameter space for $L_{dS}=1$ whose QNFs suggest a violation of SCC. We distinguish between $\omega_+$ (blue), $\omega_c$ (orange), and both (magenta).}}
\
\end{figure}

\par
\begin{figure}[t]
\centering
\includegraphics[width=0.8\linewidth]{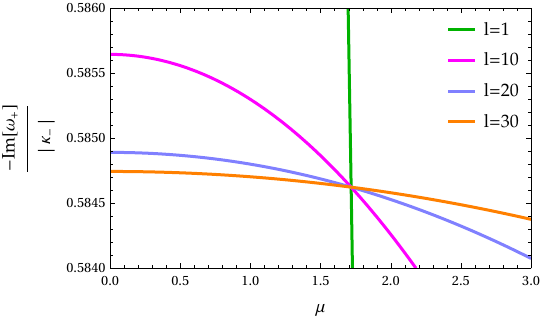}
\caption{\textit{ A graphical indication of the critical mass $\mu_{crit} \sim 1.7$ for which $\mathbb{I}m \{ \omega_+ \}/\vert \kappa_- \vert~>~1/2$. Here, $M=0.157$ and $Q=0.158$. Only in the extremal region $r_- \sim r_+$ do we observe $\mathbb{I}m \{ \omega_+ \}/\vert \kappa_- \vert $ values that approach $1/2$.}}
\label{fig:mucritSCC}
\end{figure}

 Finally, we note that for a very small parameter space on the $OU$ line corresponding to extremal black holes, we observe in Fig. \ref{fig:mucritSCC} a $\mu_{crit}$ for which $\beta > 1/2$. 

\section{Discussion and conclusion}

\par The QNM behaviour exterior to a 4D RNdS black hole has been a subject of much interest in recent years, owing in part to the instability observed in Ref. \cite{ZhuZhang2014_RNdSinstability} for the $\mu = \ell = 0$ mode in the case of a massless scalar perturbing field. Although the demonstrated stability of the scalar-, vector-, and tensor-type gravitational QNFs for $d < 6$ in Ref. \cite{refIKrn} was indicative of black hole stability against a variety of perturbing fields, the instability discovered in Ref. \cite{ZhuZhang2014_RNdSinstability} emphasised the importance of checking assumptions and exploring the full breadth of the parameter space available. This attention to detail has led to further instabilities coming to light, as in Ref. \cite{DiasReallSantos2018_SCC} for the case of small $q$ values.

\par With this in mind, a major focus of this work was to contextualise the space-time in which we perform our QNM analysis and to establish a means by which we could observe the evolution of the QNF spectrum, throughout the available phase space. This allowed for a clarification of the effect of the non-zero cosmological constant on the black hole solution space: $(i)$ the widening of the parameter space to include black hole solutions where $Q>M$, as well as analytical limitations on the possible black hole mass; $(ii)$ the presence of a cosmological horizon that confines the size of the black hole such that $r_+ \leq r_c \leq L_{dS}$; $(iii)$ the near-Nariai solution which, although unphysical, represents a black hole in which event horizon and cosmological horizon are infinitesimally close but never overlapping. 

\par With the aid of the sharkfin, the interplay between the mass, charge, and angular momentum number of the fundamental QNF with the mass and charge of the black hole can be explored fully. As first discussed in Ref. \cite{Lagos2020_Anomalous} for massive scalar QNMs in a Schwarzschild black hole space-time, there exists some critical scalar field mass $\mu_{crit}$ beyond which the magnitude of $\mathbb{I}m \{ \omega \}$ grows with increasing $\ell$; this is akin to massless scalar QNFs and therefore represents ``regular" QNF behaviour. However, for $\mu < \mu_{crit},$ $\vert \mathbb{I}m \{ \omega \} \vert$ decays with multipolar number. As noted in Ref. \cite{Papantonopoulos2022}, we observe this same behaviour for the massive, charged scalar field in the RNdS space-time. Here, we extend the analysis further. This value of $\mu_{crit}$ corresponds to a point in the QNF spectrum for which dependence on the angular momentum number is negligible (see Figs. \ref{fig:ImwVSmuNariai}, \ref{fig:ImwVSmu_MQ}, \ref{fig:ImwVSmu_MQc01}). It is interesting to note that the introduction of charge offsets this value, suggesting that the behaviour of the charged scalar field cannot be decoupled from angular momentum. The role of charge $q$ in determining $\mu_{crit}$ is therefore minimal. Similarly, black hole charge $Q$ has a negligible effect on $\mu_{crit}$: for non-zero $q$, the spacing between modes in Figs. \ref{fig:ImwVSmuNariai}, \ref{fig:ImwVSmu_MQ}, \ref{fig:ImwVSmu_MQc01} increases with $Q$. Black hole mass, on the other hand, scales inversely with $\mu_{crit}$: for $M \sim 0$, $\mu_{crit} \sim 20$ whereas for the maximum $M \sim 1/\sqrt{27}$, we obtain the minimum $\mu_{crit} \sim 1.4$. 

\par For a sense of the physical implications of this, we investigated whether there exists a possible black hole mass range within which such modes could be detected. In particular, we are interested only in modes that satisfy the FL bound. For scalar test fields oscillating within the exterior of an astrophysical black hole, recent studies \cite{BritoCardosoPani2020_Superradiance,BertiBrito2019_UltraLight1GW,BertiBrito2017a_UltraLightStochastic,BertiBrito2017b_UltraLightLIGOLISA} suggest that modes scaling as $M\mu \sim \mathcal{O}(1)$ may undergo a superradiant amplification on timescales of suitable length to be observed by current and next-generation GW detectors. Here, we find that $M \mu_{crit} \sim 0.3.$ By Eq. (\ref{eq:Mm}), we infer that only from compact objects of the order $m \lesssim 10^{-8} \; M_{\odot}$ can we expect to observe QNMs satisfying the FL bound.

\par The growing modes corresponding to superradiant amplification can be problematic for the fate of the black hole source: as demonstrated in Eq. (\ref{eq:superradiance}), superradiance can lead to instability. In the case of the RNdS black hole, Figs. \ref{fig:mu10c01l10unstable} - \ref{fig:mu10c01l10SR} demonstrate the relationship between the two phenomena: we observe that superradiant modes corresponding to $\omega_c$ occupy the same region in the parameter space as their unstable counterpart. However, we also note through Fig. \ref{fig:mu10c01l1unstable} that the semi-classical method we employ is sensitive to the values of the input parameters, with growing modes dominating the RNdS parameter space when parameters do not scale accordingly.

\par Finally, we probed the RNdS parameter space for evidence of SCC violations. Interestingly, our semi-classical analyses suggest that $\beta > 1/2$ in the near-extremal region of the RNdS black hole. In fact, this question of whether large, cold black holes respect SCC has recently been investigated in Ref. \cite{Hod2023_ColdSCC}. We also find a $\mu_{crit}$ in violation of SCC in a narrow parameter space along the $OU$ line for $Q \gtrsim M.$


\section*{Acknowledgments}

SCP was partly supported by National Research Foundation grants funded by the Korean
government (NRF-2021R1A4A2001897) and (NRF-2019R1A2C1089334). AC, ASC and AD are supported in part by the National Research Foundation (NRF) of South Africa and the SA-CERN programme; AC is also supported by a Campus France scholarship.

\appendix
\renewcommand{\theequation}{A\arabic{section}.\arabic{equation}}

\section{Details on the semi-classical method \label{app:num}}

\par We choose to make use of this method as it produces the QNF is the form of a series expansion in $L$, where black hole and scalar field input variables are left as free parameters (i.e. charges and masses do not need to be pre-defined, as is the case in a variety of other methods \cite{refKonoplyaZhidenkoReview,Dias2015_Num}). In this way, black hole and scalar field input parameters can be substituted in after the iterative procedure has been applied, therefore allowing for the interactive exploration of the RNdS phase space using the \href{https://github.com/anna-chrys/RNdS_QNMs}{interactive Mathematica notebook}.

\par For $N=n+1/2$ (where we set $n=0$), we compute the values of the QNF series terms using Eq. (\ref{eq:BorelQNF}) \cite{Hatsuda2019_WKB,Papantonopoulos2022}, for which
\begin{align}
U & =  N \sqrt{ \frac{-V^(2)}{2}} + \frac{i}{64} \left[- \frac{1}{9} \left( \frac{V^{(3)}}{V^{(2)}} \right)^2 (7 + 60N^2) + \frac{V^{(4)}}{V^{(2)}}(1+4N^2) \right] \nonumber \\
& + \frac{N}{2^{3/2} 288} \Bigg[ \frac{5}{24} \left( \frac{(V^{(3)})^4}{(-V^{(2)})^{9/2}} \right)(77+188N^2) + \frac{3}{4} \left( \frac{(V^{(3)})^2 V^{(4)}}{(-V^{(2)})^{7/2}} \right) (51+100N^2) \nonumber \\
& \quad + \frac{1}{8} \left( \frac{(V^{(4)})^2}{(-V^{(2)})^{5/2}} \right) (67+68N^2) + \left( \frac{V^{(3)} V^{(5)}}{(-V^{(2)})^{5/2}} \right) (19+28 N^2) +\left( \frac{V^{(6)}}{(-V^{(2)})^{3/2}} \right) (5+ 4N^2) \Bigg] \;. 
\end{align}

\par To solve for the critical mass, we set $\mathbb{I}m \{ \omega_{-2} \} =0$\cite{Papantonopoulos2022}. Up to $\mathcal{O}(Q^4)$, 
{\normalsize{
\begin{align}
    \mu_{crit}^2 & =  \frac{\left(3 M^2 \left(864 c^2 \left(9 B \left(90 B -17\right)+16\right)+\Lambda  \left(252 B \left(10593 B+790\right)-93491\right)\right)+19231\right)}{17496 M^2 \left(6 B+1\right) \left(126 B-11\right)} \nonumber \\
    & -\frac{ \left(8 i c Q^3 \left(9 B \left(1737 B-470\right)+115\right)\right)\sqrt{1-9 B}}{243 \sqrt{3} \left(Q^4 \left(6 B+1\right) \left(126 B-11\right)+4 M^2 Q^2 \left(234 B-11\right) \left(1-9 B\right)+120 M^4 \left(1-9 B \right)^2\right)} \nonumber \\
    & +\frac{\left(Q^2 \left(9 M^2 \left(1080 c^2 \left(9 B+2\right)-\Lambda  \left(1350 B+241\right)\right)+391\right)\right)\left(1-9 B\right)}{3645 \left(Q^4 \left(6 B+1\right) \left(126 B-11\right)+4 M^2 Q^2 \left(234 B-11\right) \left(1-9 B\right)+120 M^4 \left(1-9 B\right)^2\right)}  \nonumber \\
    & -\frac{16 i c M^2 Q \left(99 B-38\right) \left(1-9 B\right)^{3/2}}{27 \sqrt{3} \left(Q^4 \left(6 B+1\right) \left(126 B-11\right)+4 M^2 Q^2 \left(234 B-11\right) \left(1-9 B\right)+120 M^4 \left(1-9 B\right)^2\right)} \nonumber \;, \\ \label{eq:mucrit}
\end{align}
}}
\noindent where we have used $B = \Lambda M^2$ for convenience.

\bibliographystyle{JHEP} 
\bibliography{active_RNdS-QNMs}

\providecommand{\href}[2]{#2}\begingroup\raggedright\begin{thebibliography}{100}

\bibitem{Israel1966}
W.~Israel, \emph{{Singular hypersurfaces and thin shells in general
  relativity}}, \href{https://doi.org/10.1007/BF02710419}{\emph{Nuovo Cim. B}
  {\bfseries 44S10} (1966) 1}.

\bibitem{Gravitation1973}
C.W.~Misner, K.S.~Thorne and J.A.~Wheeler, \emph{{Gravitation}}, W. H. Freeman,
  San Francisco (1973).

\bibitem{Nature_BlackHoleDefinitions}
E.~Curiel, \emph{{The many definitions of a black hole}},
  \href{https://doi.org/10.1038/s41550-018-0602-1}{\emph{Nature Astron.}
  {\bfseries 3} (2019) 27}.

\bibitem{Vafa2005_Swampland}
C.~Vafa, \emph{{The String landscape and the swampland}}, .

\bibitem{ArkaniHamed2006_WGCorigins}
N.~Arkani-Hamed, L.~Motl, A.~Nicolis and C.~Vafa, \emph{{The String landscape,
  black holes and gravity as the weakest force}},
  \href{https://doi.org/10.1088/1126-6708/2007/06/060}{\emph{JHEP} {\bfseries
  06} (2007) 060}.

\bibitem{GibbonsHawking1977_BHTherm}
G.W.~Gibbons and S.W.~Hawking, \emph{{Cosmological Event Horizons,
  Thermodynamics, and Particle Creation}},
  \href{https://doi.org/10.1103/PhysRevD.15.2738}{\emph{Phys. Rev. D}
  {\bfseries 15} (1977) 2738}.

\bibitem{MossNaritaka2021_KerrdSevaporation}
R.~Gregory, I.G.~Moss, N.~Oshita and S.~Patrick, \emph{{Black hole evaporation
  in de Sitter space}},
  \href{https://doi.org/10.1088/1361-6382/ac1a68}{\emph{Class. Quant. Grav.}
  {\bfseries 38} (2021) 185005}.

\bibitem{Dolan2018_DefiningMassdS}
B.P.~Dolan, \emph{{The definition of mass in asymptotically de Sitter
  space-times}}, \href{https://doi.org/10.1088/1361-6382/ab0bdb}{\emph{Class.
  Quant. Grav.} {\bfseries 36} (2019) 077001}.

\bibitem{Anninos2012_Musings}
D.~Anninos, \emph{{De Sitter Musings}},
  \href{https://doi.org/10.1142/S0217751X1230013X}{\emph{Int. J. Mod. Phys. A}
  {\bfseries 27} (2012) 1230013}.

\bibitem{SupernovaSearchTeam1998_LCDM}
A.G.~Riess et~al., \emph{{Observational evidence from supernovae for an
  accelerating universe and a cosmological constant}},
  \href{https://doi.org/10.1086/300499}{\emph{Astron. J.} {\bfseries 116}
  (1998) 1009}.

\bibitem{SupernovaCosmologyProject1998_LCDM}
S.~Perlmutter et~al., \emph{{Measurements of $\Omega$ and $\Lambda$ from 42
  high redshift supernovae}},
  \href{https://doi.org/10.1086/307221}{\emph{Astrophys. J.} {\bfseries 517}
  (1999) 565}.

\bibitem{Planck2018}
N.~Aghanim et~al., \emph{{Planck 2018 results. VI. Cosmological parameters}},
  \href{https://doi.org/10.1051/0004-6361/201833910}{\emph{Astron. Astrophys.}
  {\bfseries 641} (2020) A6}.

\bibitem{Strominger2001_dScft}
A.~Strominger, \emph{{The dS / CFT correspondence}},
  \href{https://doi.org/10.1088/1126-6708/2001/10/034}{\emph{JHEP} {\bfseries
  10} (2001) 034}.

\bibitem{vanRiet2018_dSvacuua}
U.H.~Danielsson and T.~Van~Riet, \emph{{What if string theory has no de Sitter
  vacua?}}, \href{https://doi.org/10.1142/S0218271818300070}{\emph{Int. J. Mod.
  Phys. D} {\bfseries 27} (2018) 1830007}.

\bibitem{Dine2020_dSobstacles}
M.~Dine, J.A.P.~Law-Smith, S.~Sun, D.~Wood and Y.~Yu, \emph{{Obstacles to
  Constructing de Sitter Space in String Theory}},
  \href{https://doi.org/10.1007/JHEP02(2021)050}{\emph{JHEP} {\bfseries 02}
  (2021) 050}.

\bibitem{DolanMann2013_dStherm}
B.P.~Dolan, D.~Kastor, D.~Kubiznak, R.B.~Mann and J.~Traschen,
  \emph{{Thermodynamic Volumes and Isoperimetric Inequalities for de Sitter
  Black Holes}}, \href{https://doi.org/10.1103/PhysRevD.87.104017}{\emph{Phys.
  Rev. D} {\bfseries 87} (2013) 104017}.

\bibitem{Hawking1974_HawkT}
S.W.~Hawking, \emph{{Black hole explosions}},
  \href{https://doi.org/10.1038/248030a0}{\emph{Nature} {\bfseries 248} (1974)
  30}.

\bibitem{Hawking1975_HawkRad}
S.W.~Hawking, \emph{{Particle Creation by Black Holes}},
  \href{https://doi.org/10.1007/BF02345020}{\emph{Commun. Math. Phys.}
  {\bfseries 43} (1975) 199}.

\bibitem{vanRiet2019_FLevapBHdS}
M.~Montero, T.~Van~Riet and G.~Venken, \emph{{Festina Lente: EFT Constraints
  from Charged Black Hole Evaporation in de Sitter}},
  \href{https://doi.org/10.1007/JHEP01(2020)039}{\emph{JHEP} {\bfseries 01}
  (2020) 039}.

\bibitem{vanRiet2021_FL}
M.~Montero, C.~Vafa, T.~Van~Riet and G.~Venken, \emph{{The FL bound and its
  phenomenological implications}},
  \href{https://doi.org/10.1007/JHEP10(2021)009}{\emph{JHEP} {\bfseries 10}
  (2021) 009}.

\bibitem{SCParkDYCheong2022_FLmili}
K.~Ban, D.Y.~Cheong, H.~Okada, H.~Otsuka, J.-C.~Park and S.C.~Park,
  \emph{{Phenomenological implications on a hidden sector from the festina
  lente bound}}, \href{https://doi.org/10.1093/ptep/ptac176}{\emph{PTEP}
  {\bfseries 2023} (2023) 013B04}.

\bibitem{Lee:2021cor}
S.M.~Lee, D.Y.~Cheong, S.C.~Hyun, S.C.~Park and M.-S.~Seo, \emph{{Festina-Lente
  bound on Higgs vacuum structure and inflation}},
  \href{https://doi.org/10.1007/JHEP02(2022)100}{\emph{JHEP} {\bfseries 02}
  (2022) 100}.

\bibitem{Penrose1964_CC1}
R.~Penrose, \emph{{Gravitational collapse and space-time singularities}},
  \href{https://doi.org/10.1103/PhysRevLett.14.57}{\emph{Phys. Rev. Lett.}
  {\bfseries 14} (1965) 57}.

\bibitem{Penrose1969_CC2}
R.~Penrose, \emph{{Gravitational collapse: The role of general relativity}},
  \href{https://doi.org/10.1023/A:1016578408204}{\emph{Riv. Nuovo Cim.}
  {\bfseries 1} (1969) 252}.

\bibitem{Hawking1970_CC3}
S.W.~Hawking and R.~Penrose, \emph{{The Singularities of gravitational collapse
  and cosmology}}, \href{https://doi.org/10.1098/rspa.1970.0021}{\emph{Proc.
  Roy. Soc. Lond. A} {\bfseries 314} (1970) 529}.

\bibitem{Wald1997_WCC}
R.M.~Wald, \emph{{Gravitational collapse and cosmic censorship}},  pp.~69--85,
  10, 1997, \href{https://doi.org/10.1007/978-94-017-0934-7_5}{DOI}.

\bibitem{Chambers1997_SCC}
C.M.~Chambers, \emph{{The Cauchy horizon in black hole de sitter space-times}},
  {\emph{Annals Israel Phys. Soc.} {\bfseries 13} (1997) 33}.

\bibitem{Chandrasekhar1983}
S.~Chandrasekhar, \emph{{The Mathematical Theory of Black Holes}}, Oxford
  Univerity Press, New York (1983).

\bibitem{Penrose1999_CC}
R.~Penrose, \emph{{The question of cosmic censorship}},
  \href{https://doi.org/10.1007/BF02702355}{\emph{Journal of Astrophysics and
  Astronomy} {\bfseries 20} (1999) 233}.

\bibitem{refHorowitzBook}
G.T.~Horowitz, \emph{Black Holes in Higher Dimensions}, Cambridge University
  Press, 1~ed. (2012).

\bibitem{PoissonIsrael1990_BHinterior}
E.~Poisson and W.~Israel, \emph{{Internal structure of black holes}},
  \href{https://doi.org/10.1103/PhysRevD.41.1796}{\emph{Phys. Rev. D}
  {\bfseries 41} (1990) 1796}.

\bibitem{Dafermos2003_RNinterior}
M.~Dafermos, \emph{{The Interior of charged black holes and the problem of
  uniqueness in general relativity}}, {\emph{Commun. Pure Appl. Math.}
  {\bfseries 58} (2005) 0445}.

\bibitem{BardeenCarterHawking1973_BHmechanics}
J.M.~{Bardeen}, B.~{Carter} and S.W.~{Hawking}, \emph{{The four laws of black
  hole mechanics}},
  \href{https://doi.org/10.1007/BF01645742}{\emph{Communications in
  Mathematical Physics} {\bfseries 31} (1973) 161}.

\bibitem{Dafermos:2012np}
M.~Dafermos, \emph{{Black holes without spacelike singularities}},
  \href{https://doi.org/10.1007/s00220-014-2063-4}{\emph{Commun. Math. Phys.}
  {\bfseries 332} (2014) 729}.

\bibitem{Dafermos2018_BlueShiftLambda}
M.~Dafermos and Y.~Shlapentokh-Rothman, \emph{{Rough initial data and the
  strength of the blue-shift instability on cosmological black holes with
  $\Lambda > 0$}}, \href{https://doi.org/10.1088/1361-6382/aadbcf}{\emph{Class.
  Quant. Grav.} {\bfseries 35} (2018) 195010}.

\bibitem{Cardoso2017_QNMsSCC}
V.~Cardoso, J.a.L.~Costa, K.~Destounis, P.~Hintz and A.~Jansen,
  \emph{{Quasinormal modes and Strong Cosmic Censorship}},
  \href{https://doi.org/10.1103/PhysRevLett.120.031103}{\emph{Phys. Rev. Lett.}
  {\bfseries 120} (2018) 031103}.

\bibitem{Cardoso2018_QNMsSCC}
V.~Cardoso, J.L.~Costa, K.~Destounis, P.~Hintz and A.~Jansen, \emph{{Strong
  cosmic censorship in charged black-hole spacetimes: still subtle}},
  \href{https://doi.org/10.1103/PhysRevD.98.104007}{\emph{Phys. Rev. D}
  {\bfseries 98} (2018) 104007}.

\bibitem{Hod2018_SCC}
S.~Hod, \emph{{Strong cosmic censorship in charged black-hole spacetimes: As
  strong as ever}},
  \href{https://doi.org/10.1016/j.nuclphysb.2019.03.003}{\emph{Nucl. Phys. B}
  {\bfseries 941} (2019) 636}.

\bibitem{Mo2018_SCC}
Y.~Mo, Y.~Tian, B.~Wang, H.~Zhang and Z.~Zhong, \emph{{Strong cosmic censorship
  for the massless charged scalar field in the
  Reissner-Nordstrom\textendash{}de Sitter spacetime}},
  \href{https://doi.org/10.1103/PhysRevD.98.124025}{\emph{Phys. Rev. D}
  {\bfseries 98} (2018) 124025}.

\bibitem{DiasReallSantos2018_SCC}
O.J.C.~Dias, H.S.~Reall and J.E.~Santos, \emph{{Strong cosmic censorship for
  charged de Sitter black holes with a charged scalar field}},
  \href{https://doi.org/10.1088/1361-6382/aafcf2}{\emph{Class. Quant. Grav.}
  {\bfseries 36} (2019) 045005}.

\bibitem{DiasEperonReallSantos2018_SCC}
O.J.C.~Dias, F.C.~Eperon, H.S.~Reall and J.E.~Santos, \emph{{Strong cosmic
  censorship in de Sitter space}},
  \href{https://doi.org/10.1103/PhysRevD.97.104060}{\emph{Phys. Rev. D}
  {\bfseries 97} (2018) 104060}.

\bibitem{DiasReallSantos2018_SCC_Rough}
O.J.C.~Dias, H.S.~Reall and J.E.~Santos, \emph{{Strong cosmic censorship:
  taking the rough with the smooth}},
  \href{https://doi.org/10.1007/JHEP10(2018)001}{\emph{JHEP} {\bfseries 10}
  (2018) 001}.

\bibitem{GimGwak2019_RNdS-Lyapunov}
Y.~Gim and B.~Gwak, \emph{{Charged particle and strong cosmic censorship in
  Reissner\textendash{}Nordstr\"om\textendash{}de Sitter black holes}},
  \href{https://doi.org/10.1103/PhysRevD.100.124001}{\emph{Phys. Rev. D}
  {\bfseries 100} (2019) 124001}.

\bibitem{Konoplya2022_SCC}
R.A.~Konoplya and A.~Zhidenko, \emph{{How general is the strong cosmic
  censorship bound for quasinormal modes?}},
  \href{https://doi.org/10.1088/1475-7516/2022/11/028}{\emph{JCAP} {\bfseries
  11} (2022) 028}.

\bibitem{Hintz2015_SCC}
P.~Hintz and A.~Vasy, \emph{{Analysis of linear waves near the Cauchy horizon
  of cosmological black holes}},
  \href{https://doi.org/10.1063/1.4996575}{\emph{J. Math. Phys.} {\bfseries 58}
  (2017) 081509}.

\bibitem{Costa2016_SCC}
J.a.L.~Costa and A.T.~Franzen, \emph{{Bounded energy waves on the black hole
  interior of Reissner-Nordstr\"om-de Sitter}},
  \href{https://doi.org/10.1007/s00023-017-0592-z}{\emph{Annales Henri
  Poincare} {\bfseries 18} (2017) 3371}.

\bibitem{refRW}
T.~Regge and J.A.~Wheeler, \emph{{Stability of a Schwarzschild singularity}},
  \href{https://doi.org/10.1103/PhysRev.108.1063}{\emph{Phys. Rev.} {\bfseries
  108} (1957) 1063}.

\bibitem{refBertiCardoso}
E.~Berti, V.~Cardoso and A.O.~Starinets, \emph{{Quasinormal modes of black
  holes and black branes}},
  \href{https://doi.org/10.1088/0264-9381/26/16/163001}{\emph{Class. Quant.
  Grav.} {\bfseries 26} (2009) 163001}.

\bibitem{refKonoplyaZhidenkoReview}
R.A.~Konoplya and A.~Zhidenko, \emph{{Quasinormal modes of black holes: From
  astrophysics to string theory}},
  \href{https://doi.org/10.1103/RevModPhys.83.793}{\emph{Rev. Mod. Phys.}
  {\bfseries 83} (2011) 793}.

\bibitem{refIKrn}
H.~Kodama and A.~Ishibashi, \emph{{Master equations for perturbations of
  generalized static black holes with charge in higher dimensions}},
  \href{https://doi.org/10.1143/PTP.111.29}{\emph{Prog. Theor. Phys.}
  {\bfseries 111} (2004) 29}.

\bibitem{refIKchap6}
A.~Ishibashi and H.~Kodama, \emph{{Perturbations and Stability of Static Black
  Holes in Higher Dimensions}},
  \href{https://doi.org/10.1143/PTPS.189.165}{\emph{Prog. Theor. Phys. Suppl.}
  {\bfseries 189} (2011) 165}.

\bibitem{refKonoplya2009}
R.A.~Konoplya and A.~Zhidenko, \emph{{Instability of higher dimensional charged
  black holes in the de-Sitter world}},
  \href{https://doi.org/10.1103/PhysRevLett.103.161101}{\emph{Phys. Rev. Lett.}
  {\bfseries 103} (2009) 161101}.

\bibitem{ZhuZhang2014_RNdSinstability}
Z.~Zhu, S.-J.~Zhang, C.E.~Pellicer, B.~Wang and E.~Abdalla, \emph{{Stability of
  Reissner-Nordstr\"om black hole in de Sitter background under charged scalar
  perturbation}}, \href{https://doi.org/10.1103/PhysRevD.90.044042}{\emph{Phys.
  Rev. D} {\bfseries 90} (2014) 044042}.

\bibitem{Bekenstein1973_SR}
J.D.~Bekenstein, \emph{{Extraction of energy and charge from a black hole}},
  \href{https://doi.org/10.1103/PhysRevD.7.949}{\emph{Phys. Rev. D} {\bfseries
  7} (1973) 949}.

\bibitem{BritoCardosoPani2020_Superradiance}
R.~Brito, V.~Cardoso and P.~Pani, \emph{Superradiance}, Springer International
  Publishing (2020),
  \href{https://doi.org/10.1007/978-3-030-46622-0}{10.1007/978-3-030-46622-0}.

\bibitem{KonoplyaZhidenko2014_RNdSrprc}
R.A.~Konoplya and A.~Zhidenko, \emph{{Charged scalar field instability between
  the event and cosmological horizons}},
  \href{https://doi.org/10.1103/PhysRevD.90.064048}{\emph{Phys. Rev. D}
  {\bfseries 90} (2014) 064048}.

\bibitem{KonoplyaZhidenko2013_dRNdSinstability}
R.A.~Konoplya and A.~Zhidenko, \emph{{Instability of D-dimensional extremally
  charged Reissner-Nordstrom(-de Sitter) black holes: Extrapolation to
  arbitrary D}}, \href{https://doi.org/10.1103/PhysRevD.89.024011}{\emph{Phys.
  Rev. D} {\bfseries 89} (2014) 024011}.

\bibitem{DiasSantos2020_RNdSinstability}
O.J.C.~Dias and J.E.~Santos, \emph{{Origin of the
  Reissner-Nordstr\"om\textendash{}de Sitter instability}},
  \href{https://doi.org/10.1103/PhysRevD.102.124039}{\emph{Phys. Rev. D}
  {\bfseries 102} (2020) 124039}.

\bibitem{Papantonopoulos2022}
P.A.~Gonz\'alez, E.~Papantonopoulos, J.~Saavedra and Y.~V\'asquez,
  \emph{{Quasinormal modes for massive charged scalar fields in
  Reissner-Nordstr\"om dS black holes: anomalous decay rate}},
  \href{https://doi.org/10.1007/JHEP06(2022)150}{\emph{JHEP} {\bfseries 06}
  (2022) 150}.

\bibitem{BertiBrito2019_UltraLight1GW}
O.A.~Hannuksela, K.W.K.~Wong, R.~Brito, E.~Berti and T.G.F.~Li, \emph{{Probing
  the existence of ultralight bosons with a single gravitational-wave
  measurement}}, \href{https://doi.org/10.1038/s41550-019-0712-4}{\emph{Nature
  Astron.} {\bfseries 3} (2019) 447}.

\bibitem{BertiBrito2017b_UltraLightLIGOLISA}
R.~Brito, S.~Ghosh, E.~Barausse, E.~Berti, V.~Cardoso, I.~Dvorkin et~al.,
  \emph{{Gravitational wave searches for ultralight bosons with LIGO and
  LISA}}, \href{https://doi.org/10.1103/PhysRevD.96.064050}{\emph{Phys. Rev. D}
  {\bfseries 96} (2017) 064050}.

\bibitem{Romans1991_ColdLukewarmRNdS}
L.J.~Romans, \emph{{Supersymmetric, cold and lukewarm black holes in
  cosmological Einstein-Maxwell theory}},
  \href{https://doi.org/10.1016/0550-3213(92)90684-4}{\emph{Nucl. Phys. B}
  {\bfseries 383} (1992) 395}.

\bibitem{Mann1995_ChargedBHpairs}
R.B.~Mann and S.F.~Ross, \emph{{Cosmological production of charged black hole
  pairs}}, \href{https://doi.org/10.1103/PhysRevD.52.2254}{\emph{Phys. Rev. D}
  {\bfseries 52} (1995) 2254}.

\bibitem{Bousso1996_Nariai}
R.~Bousso, \emph{{Charged Nariai black holes with a dilaton}},
  \href{https://doi.org/10.1103/PhysRevD.55.3614}{\emph{Phys. Rev. D}
  {\bfseries 55} (1997) 3614}.

\bibitem{refCardosoDiasLemos}
V.~Cardoso, O.J.C.~Dias and J.P.S.~Lemos, \emph{{Nariai, Bertotti-Robinson and
  anti-Nariai solutions in higher dimensions}},
  \href{https://doi.org/10.1103/PhysRevD.70.024002}{\emph{Phys. Rev. D}
  {\bfseries 70} (2004) 024002}.

\bibitem{CasalsDolan2009_Nariai}
M.~Casals, S.R.~Dolan, A.C.~Ottewill and B.~Wardell, \emph{{Self-Force
  Calculations with Matched Expansions and Quasinormal Mode Sums}},
  \href{https://doi.org/10.1103/PhysRevD.79.124043}{\emph{Phys. Rev. D}
  {\bfseries 79} (2009) 124043}.

\bibitem{Belgiorno2009_chargedBHs}
F.~Belgiorno, S.L.~Cacciatori and F.~Dalla~Piazza, \emph{Pair-production of
  charged {{Dirac}} particles on charged {{Nariai}} and ultracold black hole
  manifolds}, \href{https://doi.org/10.1088/1126-6708/2009/08/028}{\emph{JHEP}
  {\bfseries 08} (2009) 028}.

\bibitem{Belgiorno2010_chargedBHs}
F.~Belgiorno, S.L.~Cacciatori and F.~Dalla~Piazza, \emph{Quantum instability
  for charged scalar particles on charged {{Nariai}} and ultracold black hole
  manifolds}, \href{https://doi.org/10.1088/0264-9381/27/5/055011}{\emph{Class.
  Quant. Grav.} {\bfseries 27} (2010) 055011}.

\bibitem{AntoniadisBenakli2020_WGCdS}
I.~Antoniadis and K.~Benakli, \emph{Weak gravity conjecture in de sitter
  space-time}, \href{https://doi.org/10.1002/prop.202000054}{\emph{Fortschritte
  der Physik} {\bfseries 68} (2020) 2000054}.

\bibitem{refNatarioSchiappa}
J.~Nat{\'a}rio and R.~Schiappa, \emph{{On the classification of asymptotic
  quasinormal frequencies for d-dimensional black holes and quantum gravity}},
  \href{https://doi.org/10.4310/ATMP.2004.v8.n6.a4}{\emph{Adv. Theor. Math.
  Phys.} {\bfseries 8} (2004) 1001}.

\bibitem{BedroyaVafa2019_TCC}
A.~Bedroya and C.~Vafa, \emph{{Trans-Planckian Censorship and the Swampland}},
  \href{https://doi.org/10.1007/JHEP09(2020)123}{\emph{JHEP} {\bfseries 09}
  (2020) 123}.

\bibitem{Kramer2003_ExactEFEsols}
H.~Stephani, D.~Kramer, M.~MacCallum, C.~Hoenselaers and E.~Herlt, \emph{Exact
  Solutions of Einstein's Field Equations}, Cambridge Monographs on
  Mathematical Physics, Cambridge University Press, 2~ed. (2003),
  \href{https://doi.org/10.1017/CBO9780511535185}{10.1017/CBO9780511535185}.

\bibitem{Bekenstein2003_BHinfo}
J.D.~Bekenstein, \emph{{Black holes and information theory}},
  \href{https://doi.org/10.1080/00107510310001632523}{\emph{Contemp. Phys.}
  {\bfseries 45} (2003) 31}.

\bibitem{refHorowitzChap1}
G.T.~Horowitz, \emph{{Black holes in four dimensions}},  in \emph{Black holes
  High. Dimens.}, G.T.~Horowitz, ed., (Cambridge), pp.~3--20, Cambridge
  University Press (2012).

\bibitem{Benakli2021_DilatonicAdS}
K.~Benakli, C.~Branchina and G.~Lafforgue-Marmet, \emph{{Dilatonic (Anti-)de
  Sitter black holes and Weak Gravity Conjecture}},
  \href{https://doi.org/10.1007/JHEP11(2021)058}{\emph{JHEP} {\bfseries 11}
  (2021) 058}.

\bibitem{MossMyers1998_CC}
P.R.~Brady, I.G.~Moss and R.C.~Myers, \emph{{Cosmic censorship: As strong as
  ever}}, \href{https://doi.org/10.1103/PhysRevLett.80.3432}{\emph{Phys. Rev.
  Lett.} {\bfseries 80} (1998) 3432}.

\bibitem{Zhang2016_ThermRNdS}
L.-C.~Zhang, R.~Zhao and M.-S.~Ma, \emph{{Entropy of
  Reissner\textendash{}Nordstr\"om\textendash{}de Sitter black hole}},
  \href{https://doi.org/10.1016/j.physletb.2016.08.013}{\emph{Phys. Lett. B}
  {\bfseries 761} (2016) 74}.

\bibitem{Brill1993_RNdSextrema}
D.R.~Brill and S.A.~Hayward, \emph{{Global structure of a black hole cosmos and
  its extremes}},
  \href{https://doi.org/10.1088/0264-9381/11/2/008}{\emph{Class. Quant. Grav.}
  {\bfseries 11} (1994) 359}.

\bibitem{Bousso1999_QuantumStructredS}
R.~Bousso, \emph{{Quantum global structure of de Sitter space}},
  \href{https://doi.org/10.1103/PhysRevD.60.063503}{\emph{Phys. Rev. D}
  {\bfseries 60} (1999) 063503}.

\bibitem{Nariai1950_StaticSol}
N.~Hidekazu, \emph{{On some static solutions of Einstein's gravitational field
  equations in a spherically symmetric case}}, {\emph{Sci. Rep. Tohoku Univ.
  Eighth Ser.} {\bfseries 34} (1950) 160}.

\bibitem{Nariai1951_NewSol}
N.~Hidekazu, \emph{{On a new cosmological solution of Einstein's field
  equations of gravitation}}, {\emph{Sci. Rep. Tohoku Univ. Eighth Ser.}
  {\bfseries 35} (1951) 46}.

\bibitem{refCardosoLemos2003}
V.~Cardoso and J.P.S.~Lemos, \emph{{Quasinormal modes of the near extremal
  Schwarzschild-de Sitter black hole}},
  \href{https://doi.org/10.1103/PhysRevD.67.084020}{\emph{Phys. Rev. D}
  {\bfseries 67} (2003) 084020}.

\bibitem{GinspargPerry1982_NariaidS}
P.H.~Ginsparg and M.J.~Perry, \emph{{Semiclassical Perdurance of de Sitter
  Space}}, \href{https://doi.org/10.1016/0550-3213(83)90636-3}{\emph{Nucl.
  Phys. B} {\bfseries 222} (1983) 245}.

\bibitem{Cardoso2003_ExtremalSchw}
V.~Cardoso and J.P.S.~Lemos, \emph{{Quasinormal modes of the near extremal
  Schwarzschild-de Sitter black hole}},
  \href{https://doi.org/10.1103/PhysRevD.67.084020}{\emph{Phys. Rev. D}
  {\bfseries 67} (2003) 084020}.

\bibitem{Molina2003_ExtremalDdimBHs}
C.~Molina, \emph{{Quasinormal modes of d-dimensional spherical black holes with
  near extreme cosmological constant}},
  \href{https://doi.org/10.1103/PhysRevD.68.064007}{\emph{Phys. Rev. D}
  {\bfseries 68} (2003) 064007}.

\bibitem{Churilova2021_ExtremalAnal}
M.S.~Churilova, R.A.~Konoplya and A.~Zhidenko, \emph{{Analytic formula for
  quasinormal modes in the near-extreme Kerr-Newman\textendash{}de Sitter
  spacetime governed by a non-P\"oschl-Teller potential}},
  \href{https://doi.org/10.1103/PhysRevD.105.084003}{\emph{Phys. Rev. D}
  {\bfseries 105} (2022) 084003}.

\bibitem{Bertotti1959_BRsol}
B.~Bertotti, \emph{{Uniform electromagnetic field in the theory of general
  relativity}}, \href{https://doi.org/10.1103/PhysRev.116.1331}{\emph{Phys.
  Rev.} {\bfseries 116} (1959) 1331}.

\bibitem{Robinson1959_BRsol}
I.~Robinson, \emph{{A Solution of the Maxwell-Einstein Equations}},
  {\emph{Bull. Acad. Pol. Sci. Ser. Sci. Math. Astron. Phys.} {\bfseries 7}
  (1959) 351}.

\bibitem{HawkingRoss1995_NariaiRNdS}
S.W.~Hawking and S.F.~Ross, \emph{{Duality between electric and magnetic black
  holes}}, \href{https://doi.org/10.1103/PhysRevD.52.5865}{\emph{Phys. Rev. D}
  {\bfseries 52} (1995) 5865}.

\bibitem{BoussoHawking1996_PairCreation}
R.~Bousso and S.W.~Hawking, \emph{{Pair creation of black holes during
  inflation}}, \href{https://doi.org/10.1103/PhysRevD.54.6312}{\emph{Phys. Rev.
  D} {\bfseries 54} (1996) 6312}.

\bibitem{Vishveshwara1970}
C.V.~Vishveshwara, \emph{Scattering of gravitational radiation by a
  schwarzschild black-hole},
  \href{https://doi.org/10.1038/227936a0}{\emph{Nature} {\bfseries 227} (1970)
  936}.

\bibitem{Press1971}
W.H.~Press, \emph{{Long Wave Trains of Gravitational Waves from a Vibrating
  Black Hole}}, \href{https://doi.org/10.1086/180849}{\emph{Astrophys. J.
  Lett.} {\bfseries 170} (1971) L105}.

\bibitem{Echeverria1989_BHpropertiesestimate}
F.~Echeverria, \emph{Gravitational-wave measurements of the mass and angular
  momentum of a black hole},
  \href{https://doi.org/10.1103/PhysRevD.40.3194}{\emph{Phys. Rev. D}
  {\bfseries 40} (1989) 3194}.

\bibitem{LIGO2019_GWTC1-GRtest}
B.P.~Abbott et~al., \emph{{Tests of General Relativity with the Binary Black
  Hole Signals from the LIGO-Virgo Catalog GWTC-1}},
  \href{https://doi.org/10.1103/PhysRevD.100.104036}{\emph{Phys. Rev. D}
  {\bfseries 100} (2019) 104036}.

\bibitem{LIGO2020_GWTC2-GRtest_pyRing3}
R.~Abbott et~al., \emph{{Tests of general relativity with binary black holes
  from the second LIGO-Virgo gravitational-wave transient catalog}},
  \href{https://doi.org/10.1103/PhysRevD.103.122002}{\emph{Phys. Rev. D}
  {\bfseries 103} (2021) 122002}.

\bibitem{LIGO2021_GWTC3-GRtest}
R.~Abbott et~al., \emph{{Tests of General Relativity with GWTC-3}}, .

\bibitem{Carullo2019_pyRing1}
G.~Carullo, W.~Del~Pozzo and J.~Veitch, \emph{{Observational Black Hole
  Spectroscopy: A time-domain multimode analysis of GW150914}},
  \href{https://doi.org/10.1103/PhysRevD.99.123029}{\emph{Phys. Rev. D}
  {\bfseries 99} (2019) 123029}.

\bibitem{refNoHair_pyRing2}
M.~Isi, M.~Giesler, W.M.~Farr, M.A.~Scheel and S.A.~Teukolsky, \emph{{Testing
  the no-hair theorem with GW150914}},
  \href{https://doi.org/10.1103/PhysRevLett.123.111102}{\emph{Phys. Rev. Lett.}
  {\bfseries 123} (2019) 111102}.

\bibitem{Stairs2003_GRpulsarTests}
I.H.~Stairs, \emph{{Testing general relativity with pulsar timing}},
  \href{https://doi.org/10.12942/lrr-2003-5}{\emph{Living Rev. Rel.} {\bfseries
  6} (2003) 5}.

\bibitem{Will2009_GRstellarTests}
D.~Merritt, T.~Alexander, S.~Mikkola and C.M.~Will, \emph{{Testing Properties
  of the Galactic Center Black Hole Using Stellar Orbits}},
  \href{https://doi.org/10.1103/PhysRevD.81.062002}{\emph{Phys. Rev. D}
  {\bfseries 81} (2010) 062002}.

\bibitem{refWillReview2014}
C.M.~Will, \emph{{The confrontation between general relativity and
  experiment}}, \href{https://doi.org/10.12942/lrr-2014-4}{\emph{Living Rev.
  Rel.} {\bfseries 17} (2014) 4}.

\bibitem{Berti2015_TestingGRastro}
E.~Berti et~al., \emph{{Testing General Relativity with Present and Future
  Astrophysical Observations}},
  \href{https://doi.org/10.1088/0264-9381/32/24/243001}{\emph{Class. Quant.
  Grav.} {\bfseries 32} (2015) 243001}.

\bibitem{refNollert1999}
H.-P.~Nollert, \emph{{Quasinormal modes: the characteristic `sound' of black
  holes and neutron stars}},
  \href{https://doi.org/10.1088/0264-9381/16/12/201}{\emph{Class. Quant. Grav.}
  {\bfseries 16} (1999) R159}.

\bibitem{refKokkotasRev}
K.D.~Kokkotas and B.G.~Schmidt, \emph{{Quasinormal modes of stars and black
  holes}}, \href{https://doi.org/10.12942/lrr-1999-2}{\emph{Living Rev. Rel.}
  {\bfseries 2} (1999) 2}.

\bibitem{refFerrari2008}
V.~Ferrari and L.~Gualtieri, \emph{{Quasi-Normal Modes and Gravitational Wave
  Astronomy}}, \href{https://doi.org/10.1007/s10714-007-0585-1}{\emph{Gen. Rel.
  Grav.} {\bfseries 40} (2008) 945}.

\bibitem{Grandclement2007_Spectral}
P.~Grandclement and J.~Novak, \emph{{Spectral methods for numerical
  relativity}}, \href{https://doi.org/10.12942/lrr-2009-1}{\emph{Living Rev.
  Rel.} {\bfseries 12} (2009) 1}.

\bibitem{Dias2015_Num}
O.J.C.~Dias, J.E.~Santos and B.~Way, \emph{{Numerical Methods for Finding
  Stationary Gravitational Solutions}},
  \href{https://doi.org/10.1088/0264-9381/33/13/133001}{\emph{Class. Quant.
  Grav.} {\bfseries 33} (2016) 133001}.

\bibitem{refZerilli}
F.J.~Zerilli, \emph{{Gravitational field of a particle falling in a
  schwarzschild geometry analyzed in tensor harmonics}},
  \href{https://doi.org/10.1103/PhysRevD.2.2141}{\emph{Phys. Rev. D} {\bfseries
  2} (1970) 2141}.

\bibitem{Hod2018_RNdS}
S.~Hod, \emph{{The instability spectra of near-extremal Reissner-Nordstr\"om-de
  Sitter black holes}},
  \href{https://doi.org/10.1016/j.physletb.2018.09.039}{\emph{Phys. Lett. B}
  {\bfseries 786} (2018) 217}.

\bibitem{DavisPrice1971}
M.~Davis, R.~Ruffini, W.H.~Press and R.H.~Price, \emph{{Gravitational radiation
  from a particle falling radially into a Schwarzschild black hole}},
  \href{https://doi.org/10.1103/PhysRevLett.27.1466}{\emph{Phys. Rev. Lett.}
  {\bfseries 27} (1971) 1466}.

\bibitem{Price1994}
C.~Gundlach, R.H.~Price and J.~Pullin, \emph{{Late time behavior of stellar
  collapse and explosions: 1. Linearized perturbations}},
  \href{https://doi.org/10.1103/PhysRevD.49.883}{\emph{Phys. Rev. D} {\bfseries
  49} (1994) 883}.

\bibitem{refLeaver1985}
E.W.~Leaver, \emph{{An analytic representation for the quasi normal modes of
  Kerr black holes}}, \href{https://doi.org/10.1098/rspa.1985.0119}{\emph{Proc.
  Roy. Soc. Lond. A} {\bfseries 402} (1985) 285}.

\bibitem{PoschlTellerPotential}
G.~P{\"o}schl and E.~Teller, \emph{{Bemerkungen zur Quantenmechanik des
  anharmonischen Oszillators}},
  \href{https://doi.org/10.1007/BF01331132}{\emph{Z. Phys.} {\bfseries 83}
  (1933) 143}.

\bibitem{PoschlTellerMethod}
H.-J.~Blome and B.~Mashhoon, \emph{Quasi-normal oscillations of a schwarzschild
  black hole},
  \href{https://doi.org/https://doi.org/10.1016/0375-9601(84)90769-2}{\emph{Physics
  Letters A} {\bfseries 100} (1984) 231}.

\bibitem{Dias2009_Paraspectral1}
O.J.C.~Dias, P.~Figueras, R.~Monteiro, J.E.~Santos and R.~Emparan,
  \emph{{Instability and new phases of higher-dimensional rotating black
  holes}}, \href{https://doi.org/10.1103/PhysRevD.80.111701}{\emph{Phys. Rev.
  D} {\bfseries 80} (2009) 111701}.

\bibitem{Dias2009_Paraspectral2}
O.J.C.~Dias, P.~Figueras, R.~Monteiro, H.S.~Reall and J.E.~Santos, \emph{{An
  instability of higher-dimensional rotating black holes}},
  \href{https://doi.org/10.1007/JHEP05(2010)076}{\emph{JHEP} {\bfseries 05}
  (2010) 076}.

\bibitem{refAIM_OG}
H.~Ciftci, R.L.~Hall and N.~Saad, \emph{{Asymptotic iteration method for
  eigenvalue problems}},
  \href{https://doi.org/10.1088/0305-4470/36/47/008}{\emph{J. Phys. A. Math.
  Gen.} {\bfseries 36} (2003) 11807}.

\bibitem{refAIM}
H.T.~Cho, A.S.~Cornell, J.~Doukas and W.~Naylor, \emph{{Black hole quasinormal
  modes using the asymptotic iteration method}},
  \href{https://doi.org/10.1088/0264-9381/27/15/155004}{\emph{Class. Quant.
  Grav.} {\bfseries 27} (2010) 155004}.

\bibitem{HorowitzHubeny}
G.T.~Horowitz and V.E.~Hubeny, \emph{{Quasinormal modes of AdS black holes and
  the approach to thermal equilibrium}},
  \href{https://doi.org/10.1103/PhysRevD.62.024027}{\emph{Phys. Rev. D}
  {\bfseries 62} (2000) 024027}.

\bibitem{refBHWKB0}
B.F.~Schutz and C.M.~Will, \emph{{Black hole normal modes - A semianalytic
  approach}}, \href{https://doi.org/10.1086/184453}{\emph{Astrophys. J.}
  {\bfseries 291} (1985) L33}.

\bibitem{refBHWKB0.5}
C.M.~Will, \emph{{Approximation methods in gravitational-radiation theory}},
  \href{https://doi.org/10.1139/p86-023}{\emph{Can. J. Phys.} {\bfseries 64}
  (1986) 140}.

\bibitem{refBHWKB1}
S.~Iyer and C.M.~Will, \emph{{Black-hole normal modes: A WKB approach. I.
  Foundations and application of a higher-order WKB analysis of
  potential-barrier scattering}},
  \href{https://doi.org/10.1103/PhysRevD.35.3621}{\emph{Phys. Rev. D}
  {\bfseries 35} (1987) 3621}.

\bibitem{Konoplya2003}
R.A.~Konoplya, \emph{{Quasinormal behavior of the d-dimensional Schwarzschild
  black hole and higher order WKB approach}},
  \href{https://doi.org/10.1103/PhysRevD.68.024018}{\emph{Phys. Rev. D}
  {\bfseries 68} (2003) 024018}.

\bibitem{Konoplya2019}
R.A.~Konoplya, A.~Zhidenko and A.F.~Zinhailo, \emph{{Higher order WKB formula
  for quasinormal modes and grey-body factors: recipes for quick and accurate
  calculations}}, \href{https://doi.org/10.1088/1361-6382/ab2e25}{\emph{Class.
  Quant. Grav.} {\bfseries 36} (2019) 155002}.

\bibitem{refDolanOttewill2009}
S.R.~Dolan and A.C.~Ottewill, \emph{{On an expansion method for black hole
  quasinormal modes and Regge poles}},
  \href{https://doi.org/10.1088/0264-9381/26/22/225003}{\emph{Class. Quant.
  Grav.} {\bfseries 26} (2009) 225003}.

\bibitem{refGoebel1972}
C.J.~Goebel, \emph{Comments on the \enquote{vibrations} of a black hole},
  {\emph{Astrophys. J.} {\bfseries 172} (1972) }.

\bibitem{refOurLargeL}
C.-H.~Chen, H.-T.~Cho, A.~Chrysostomou and A.S.~Cornell, \emph{Quasinormal
  modes for integer and half-integer spins within the large angular momentum
  limit}, \href{https://doi.org/10.1103/PhysRevD.104.024009}{\emph{Phys. Rev.
  D} {\bfseries 104} (2021) 024009}.

\bibitem{Vishveshwara1970stable}
C.V.~Vishveshwara, \emph{{Stability of the schwarzschild metric}},
  \href{https://doi.org/10.1103/PhysRevD.1.2870}{\emph{Phys. Rev. D} {\bfseries
  1} (1970) 2870}.

\bibitem{Dolan2010_KerrQNMs}
S.R.~Dolan, \emph{{The Quasinormal Mode Spectrum of a Kerr Black Hole in the
  Eikonal Limit}},
  \href{https://doi.org/10.1103/PhysRevD.82.104003}{\emph{Phys. Rev. D}
  {\bfseries 82} (2010) 104003}.

\bibitem{BenderWu1969_AnharmonicOscillator}
C.M.~Bender and T.T.~Wu, \emph{Anharmonic oscillator},
  \href{https://doi.org/10.1103/PhysRev.184.1231}{\emph{Phys. Rev.} {\bfseries
  184} (1969) 1231}.

\bibitem{refFerrMashh1}
V.~Ferrari and B.~Mashhoon, \emph{{New approach to the quasinormal modes of a
  black hole}}, \href{https://doi.org/10.1103/PhysRevD.30.295}{\emph{Phys. Rev.
  D} {\bfseries 30} (1984) 295}.

\bibitem{refFerrMashh2}
V.~Ferrari and B.~Mashhoon, \emph{{Oscillations of a Black Hole}},
  \href{https://doi.org/10.1103/PhysRevLett.52.1361}{\emph{Phys. Rev. Lett.}
  {\bfseries 52} (1984) 1361}.

\bibitem{Hatsuda2019_WKB}
Y.~Hatsuda, \emph{{Quasinormal modes of black holes and Borel summation}},
  \href{https://doi.org/10.1103/PhysRevD.101.024008}{\emph{Phys. Rev. D}
  {\bfseries 101} (2020) 024008}.

\bibitem{KonoplyaZhidenko2022_Bernstein}
R.A.~Konoplya and A.~Zhidenko, \emph{{Bernstein spectral method for quasinormal
  modes of a generic black hole spacetime and application to instability of
  dilaton-de Sitter solution}}, .

\bibitem{Lagos2020_Anomalous}
M.~Lagos, P.G.~Ferreira and O.J.~Tattersall, \emph{{Anomalous decay rate of
  quasinormal modes}},
  \href{https://doi.org/10.1103/PhysRevD.101.084018}{\emph{Phys. Rev. D}
  {\bfseries 101} (2020) 084018}.

\bibitem{Starobinsky1973_SR1}
A.A.~Starobinsky, \emph{{Amplification of waves reflected from a rotating
  ''black hole''.}}, {\emph{Sov. Phys. JETP} {\bfseries 37} (1973) 28}.

\bibitem{Starobinsky1974_SR2}
A.A.~Starobinskil and S.M.~Churilov, \emph{{Amplification of electromagnetic
  and gravitational waves scattered by a rotating ''black hole''}}, {\emph{Sov.
  Phys. JETP} {\bfseries 65} (1974) 1}.

\bibitem{BertiBrito2017a_UltraLightStochastic}
R.~Brito, S.~Ghosh, E.~Barausse, E.~Berti, V.~Cardoso, I.~Dvorkin et~al.,
  \emph{{Stochastic and resolvable gravitational waves from ultralight
  bosons}}, \href{https://doi.org/10.1103/PhysRevLett.119.131101}{\emph{Phys.
  Rev. Lett.} {\bfseries 119} (2017) 131101}.

\bibitem{Arvanitaki2009_StringAxiverse}
A.~Arvanitaki, S.~Dimopoulos, S.~Dubovsky, N.~Kaloper and J.~March-Russell,
  \emph{{String Axiverse}},
  \href{https://doi.org/10.1103/PhysRevD.81.123530}{\emph{Phys. Rev. D}
  {\bfseries 81} (2010) 123530}.

\bibitem{PDG2022}
R.L.~Workman and Others, \emph{{Review of Particle Physics}},
  \href{https://doi.org/10.1093/ptep/ptac097}{\emph{PTEP} {\bfseries 2022}
  (2022) 083C01}.

\bibitem{Chrysostomou2022_BHnilmanifold}
A.~Chrysostomou, A.~Cornell, A.~Deandrea, E.~Ligout and D.~Tsimpis,
  \emph{{Black holes and nilmanifolds: quasinormal modes as the fingerprints of
  extra dimensions?}},
  \href{https://doi.org/10.1140/epjc/s10052-023-11496-w}{\emph{Eur. Phys. J. C}
  {\bfseries 83} (2023) 325}.

\bibitem{Hod2023_ColdSCC}
S.~Hod, \emph{{Black holes that are too cold to respect cosmic censorship}}, .

\end{thebibliography}\endgroup

\end{document}